\documentclass[12pt, a4paper]{article}
\pdfoutput=1
\usepackage[height=25cm,width=16cm]{geometry}
\usepackage{amsfonts}
\usepackage{amsmath}
\usepackage{amssymb}
\usepackage{ascmac}
\usepackage{dcolumn}
\usepackage{bm,here}
\usepackage{subfig}
\usepackage{comment}
\usepackage{ifpdf}
\usepackage{slashed}
\usepackage{colortbl}
\usepackage{color}
\usepackage[mathscr]{eucal}
\usepackage[sort&compress, numbers, merge]{natbib}

\usepackage{cancel}

\ifpdf
  \usepackage{graphicx}     
  \usepackage[bookmarksopen,colorlinks=true,linkcolor=bblue,citecolor=ppink,urlcolor=ppink]{hyperref}
\else     
  \usepackage[dvipdfmx]{graphicx}     
  \usepackage[dvipdfmx,bookmarksopen,colorlinks=true,linkcolor=bblue,citecolor=ppink,urlcolor=ppink]{hyperref}
\fi

\usepackage{multicol}
\definecolor{red}{rgb}{1,0,0}
\definecolor{ppink}{rgb}{0.921545,0.440586,0.687243}
\definecolor{bblue}{rgb}{0.400000,0.400000,1.000000}
\usepackage[charter]{mathdesign}

\begin{document}

\begin{titlepage}

\begin{flushright}
\hfill IPMU16-0045 \\
\hfill PITT-PACC 1517\\
\end{flushright}

\begin{center}

\vskip 0.5cm
{\Large \bf Effective Theory of WIMP Dark Matter \\[.3em] supplemented by Simplified Models:}\\[0.7em]
{\large \bf Singlet-like Majorana fermion case}

\vskip 1.5cm
{\large
Shigeki Matsumoto$^{(a)}$,
Satyanarayan Mukhopadhyay$^{\,(a, b)}$,\\[.5em]
and Yue-Lin Sming Tsai$^{(a)}$
}

\vskip 1.0cm
\begin{tabular}{l}
$^{(a)}$ {\sl Kavli IPMU (WPI), UTIAS, University of Tokyo, Kashiwa, 277-8583, Japan}\\[.3em]
$^{(b)}$ {\sl PITT-PACC, Department of Physics and Astronomy, University of Pittsburgh,}\\[.1em]
\hspace{28ex} {\sl PA 15260, USA}
\end{tabular}

\vskip 1.5cm
\begin{abstract}
\noindent
We enumerate the set of simplified models which match onto the complete set of gauge invariant effective operators up to dimension six describing interactions of a singlet-like Majorana fermion dark matter with the standard model. Tree level matching conditions for each case are worked out in the large mediator mass limit, defining a one to one correspondence between the effective operator coefficients and the simplified model parameters for weakly interacting models. Utilizing such a mapping, we compute the dark matter annihilation rate in the early universe, as well as other low-energy observables like nuclear recoil rates using the effective operators, while the simplified models are used to compute the dark matter production rates at high energy colliders like LEP, LHC and future lepton colliders. Combining all relevant constraints with a profile likelihood analysis, we then discuss the currently allowed parameter regions and prospects for future searches in terms of the effective operator parameters, reducing the model dependence to a minimal level. In the parameter region where such a model-independent analysis is applicable, and leaving aside the special dark matter mass regions where the annihilation proceeds through an s-channel $Z$ or Higgs boson pole, the current constraints allow effective operator suppression scales ($\Lambda$) of the order of a few hundred GeV for dark matter masses $m_\chi >$ 20\,GeV at $95\%$ C.L., while the maximum allowed scale is around $3$ TeV for $m_\chi \sim \mathcal{O}(1\,{\rm TeV})$. An estimate of the future reach of ton-scale direct detection experiments and planned electron-positron colliders show that most of the remaining regions can be probed, apart from dark matter masses near half of the Z-boson mass (with $500\,{\rm GeV} < \Lambda < 2\,{\rm TeV} $) and those beyond the kinematic reach of the future lepton colliders.
\end{abstract}

\end{center}

\end{titlepage}

\tableofcontents
\newpage
\setcounter{page}{1}

\section{Introduction}
\label{sec: intro}

Dark matter candidates charged under the weak isospin can interact with the standard model (SM) particles via known gauge interactions, a fact that simplifies their phenomenology considerably. A gauge singlet scalar field can also have a renormalizable interaction with the SM Higgs doublet. However, in order to couple the SM sector to a gauge singlet fermion dark matter (stabilized by a postulated $Z_2$ symmetry), one needs to introduce either additional bosonic degrees of freedom that couple to both the sectors, or additional fermionic degrees of freedom with electroweak charges which can mix with the singlet state after electroweak symmetry breaking. Possible frameworks to discuss the phenomenology of a singlet-like fermion WIMP (weakly interacting massive particle) candidate have been studied at length for decades, from specific ultra-violet (UV) complete models (e.g., the bino in the MSSM), to model independent setups with effective operators.

The three most relevant observables for a stable WIMP are it's relic density obtained via thermal freeze out, it's pair annihilation cross-section in the current epoch in dark matter dense regions, and the elastic scattering rate of WIMP's with nuclei. All of these processes involve scatterings with an energy scale or momentum transfer comparable to the WIMP mass or much lower. Therefore, as long as the new states mediating the interactions are at least a few times heavier than the dark matter, we can work with a set of effective operators of leading dimension relevant to the process, and truncating the operator series would not lead to any inconsistency nor amount to large theoretical errors. 

Pair production of WIMP's are being searched for at high-energy colliders as well, and for certain types of interactions or for lower dark matter masses, they could be competitive with, or even have a larger reach than that of the direct or indirect detection probes. Collider searches for dark matter necessarily rely on a recoil of the dark matter pair against a hard radiation in the initial state, which can lead to a sub-process centre of mass energy ($\hat{s}^{1/2}$) considerably higher than the dark matter pair mass. Therefore, in such a situation, the effective operator description would be accurate in the region where the suppression scale of the operators are higher than the $\hat{s}^{1/2}$ involved, or we should discard events from the analysis with $\hat{s}^{1/2}$ larger than the suppression scale of the effective field theory (EFT)\,\cite{Racco:2015dxa}. An alternative approach would be to use simplified models including both the dark matter and the mediators as new states to interpret the collider searches, in which case inevitably the number of possibilities to study is large, and for comparison with other constraints on dark matter, we need to perform multiple sets of global analyses. On the other hand, if we focus on the parameter region where we can be model-independent as far as relic density, direct and indirect detections are concerned by using effective operators to describe these processes, the collider searches can be interpreted in terms of simplified models which match onto those effective operators, in the limit of very heavy mediators.

We propose to carry out such a programme for the singlet Majorana fermion dark matter in this study, the end goal being performing a complete likelihood analysis with a proper treatment of all data at hand. Once a one-to-one correspondence between the simplified model parameters and the EFT parameters are established, we can consistently compute the collider cross-sections within the simplified models, and rest of the observables with the corresponding effective operators. The final results will be presented in terms of the EFT parameters. Such an approach does not require any additional effort from the experimental collaborations of high energy colliders, an upper bound on the relevant cross-sections after a set of kinematical cuts can be interpreted within this framework without losing any information from the tail of the missing momentum distributions. At the same time, interpreting the results within the EFT parameters helps us create a global picture of WIMP's without any prejudice to a particular new physics model.

The SM gauge invariant set of effective operators that describe the interactions of a Majorana fermion WIMP are well studied, and we provide a brief review on the operators in Sec.\,\ref{sec: EFT}, following our previous study in Ref.\,\cite{Matsumoto:2014rxa}. Therefore, the first step in our approach is to write down all possible simplified models that would match onto each of the effective operators when the mediating particles are heavy enough. We carry out this exercise (parts of which have been reported in Ref.\,\cite{Dreiner:2012xm}) in Sec.\,\ref{sec: simplified models}. The first half of Sec.\,\ref{sec: application} is devoted to a discussion of all the relevant experimental constraints and the construction of the likelihood function for each. In the second half of this section we first address the question of direct search for the mediators versus the monojet plus missing momentum search at the LHC, and then go on to discuss the results of our likelihood analysis, where the latter include only the monojet constraints for reasons explained in that section. We summarize our study in Sec.\,\ref{sec: summary}, with a view towards the role of future experiments in probing the parameter space that survive all current constraints. 

\section{Effective operators in dark matter interactions}
\label{sec: EFT}

In this section, we briefly review the effective field theory of a Majorana fermion dark matter (DM) in which the low-energy degrees of freedom consist of the SM fields and the DM field $\chi$\,\cite{Matsumoto:2014rxa}. The EFT is described by the following Lagrangian in general:
\begin{eqnarray}
	{\cal L}_{\rm EFT} =
	{\cal L}_{\rm SM} +
	\frac{1}{2}\bar{\chi}(i\slashed{\partial} - M_\chi)\chi +
	\frac{1}{2}\sum_{a,\,n} c_a \frac{{\cal O}_a} {\Lambda^{n - 4}_a},
	\label{eq: EFT}
\end{eqnarray}
where ${\cal L}_{\rm SM}$ is the renormalizable SM Lagrangian, $M_\chi$ is the bare DM mass parameter before the electroweak symmetry breaking, and ${\cal O}_a$ represent operators of mass dimension $n$ describing interactions between the DM and the SM particles, with dimensionless coefficients $c_a$. We assume that physics behind the EFT is described by a weakly interacting theory and restrict $|c_a|<1$. For such theories, the suppression scale $\Lambda_a$ for a tree generated operator corresponds to the mass of a heavy intermediate particle generating ${\cal O}_a$.

The set of operators ${\cal O}_a$ are required to respect the SM gauge symmetry, and an additional $Z_2$ symmetry to ensure the stability of the DM particle, under which the DM field is odd while the SM fields are even. If the DM field $\chi$ is a singlet under the SM gauge symmetry, we have eight types of independent operators up to mass dimension-six; all of which are shown in Tab.\,\ref{tab: operators} with $H$, $Q_i$, $U_i$, $D_i$, $L_i$ and $E_i$ denoting the Higgs doublet, the quark doublets, the up-type quark singlets, the down-type quark singlets, the lepton doublets and the charged lepton singlets, respectively.\footnote{In our notation, an appropriate chirality projection operator $P_R$ or $P_L$ is associated with an SM fermion field, e.g., $\bar{Q}_i \gamma_\mu Q_j$ is equivalent to $\bar{Q}_i \gamma_\mu P_L Q_j$, while $\bar{U}_i \gamma_\mu U_j$ is the same as $\bar{U}_i \gamma_\mu P_R U_j$.} Here, the subscripts `$i$' and `$j$' represent flavour indices, while $D_\mu$ is the covariant derivative acting on the Higgs field $H$. Some comments on the SM gauge invariant operators in this table are in order below:
\begin{itemize}
	\setlength{\itemsep}{0mm}

	\item
	The operators constitute a complete set up to mass dimension six, so that other operators such as the anapole moment, $(\bar{\chi} \gamma_\mu \gamma_5 \chi)\,\partial^\mu B_{\mu\nu}$, with $B_{\mu\nu}$ being the field strength tensor of the hypercharge gauge boson, can be written as a linear combination of this operator basis using equations of motion.

	\item
	All interactions between the DM and the SM particles are from higher dimensional operators because of the $Z_2$ symmetry. Furthermore, all except ${\cal O}_{PS}$ preserve the CP symmetry.

	\item
	Since the DM is a Majorana fermion, a factor of $1/2$ is included in front of the interaction terms, following standard normalization. 

	\item
	The DM particle is not necessarily a pure singlet under the SM gauge symmetry, though the DM field $\chi$ is a singlet. In fact, $\chi$ can mix with another $Z_2$ odd SU(2)$_L$ doublet field (which can be introduced in a simplified model) after the electroweak symmetry breaking, and it generates ${\cal O}_S$, ${\cal O}_{PS}$ and ${\cal O}_H$, as shown in the next section.

	\item
	There is an implicit assumption that all suppression scales $\Lambda_a$ are of a similar order. We also assume that mass dimension 7 operators play a sub-leading role.

	\item
	Generically, one needs to fix a scale defining the EFT Lagrangian\,(\ref{eq: EFT}), and consider RGE effects on the Wilson coefficients $c_a$ to calculate physical quantities\,\cite{Georgi:1994qn}. Since we are assuming a weakly interacting theory behind the EFT, these effects do not give a sizeable contribution. We therefore neglect the RGE effects for simplicity, without inducing large errors for weakly coupled UV theories.

\end{itemize}

\begin{table}[t]
	\begin{center}
	\begin{tabular}{| c l l l |}
	\hline
	Dim.\,5 &
	${\cal O}_S = (\bar{\chi} \chi) |H|^2$ &
	${\cal O}_{PS} = (\bar{\chi} i\gamma_5 \chi) |H|^2$ & \\
	Dim.\,6 &
	${\cal O}_Q = (\bar{\chi} \gamma^\mu \gamma_5 \chi)(\bar{Q}_i \gamma_\mu Q_j)$ &
	${\cal O}_U = (\bar{\chi} \gamma^\mu \gamma_5 \chi)(\bar{U}_i \gamma_\mu U_j)$ & 
	${\cal O}_D = (\bar{\chi} \gamma^\mu \gamma_5 \chi)(\bar{D}_i \gamma_\mu D_j)$ \\
	&
	${\cal O}_L = (\bar{\chi} \gamma^\mu \gamma_5 \chi)(\bar{L}_i \gamma_\mu L_j)$ &
	${\cal O}_E = (\bar{\chi} \gamma^\mu \gamma_5 \chi)(\bar{E}_i \gamma_\mu E_j)$ &
	${\cal O}_H = (\bar{\chi} \gamma^\mu \gamma_5 \chi)(H^\dagger i\overleftrightarrow{D_\mu} H)$ \\
	\hline
	\end{tabular}
	\caption{\sl\small SM gauge invariant operators for interactions between the DM and the SM particles.}
	\label{tab: operators}
	\end{center}
\end{table}

It is convenient to rewrite the Lagrangian\,(\ref{eq: EFT}) using the lowest suppression scale $\Lambda \equiv {\rm min}(\Lambda_a)$, which makes our numerical analysis easier. The Lagrangian then reads
\begin{eqnarray}
	{\cal L}_{\rm EFT} =
	{\cal L}_{\rm SM} +
	\frac{1}{2}\bar{\chi}(i\slashed{\partial} - M_\chi)\chi +
	\frac{1}{2}\sum_{a,\,n} \tilde{c}_a \frac{{\cal O}_a} {\Lambda^{n - 4}},
	\label{eq: EFT2}
\end{eqnarray}
where the coefficients $\tilde{c}_a$ are defined by $\tilde{c}_a \equiv c_a (\Lambda/\Lambda_a)^{(n - 4)}$. Since the absolute value of $\tilde{c}_a$ is always smaller than that of $c_a$, the complete model parameter space defined by the original effective Lagrangian\,(\ref{eq: EFT}) is covered by varying all $\tilde{c}_a$'s in the region $|\tilde{c}_a| \lesssim 1$.

\section{Connection between the EFT and simplified models}
\label{sec: simplified models}

When the physical mass of the DM (denoted by $m_{\rm DM}$) and the electroweak scale are much lower than the suppression scales $\Lambda_a$, the EFT can be used to describe non-relativistic DM phenomena such as DM freeze-out in the early universe and DM annihilation in the present universe or its scattering with nuclei. On the other hand, when we consider relativistic DM phenomena such as DM production at the LHC, the EFT description is of limited applicability since a clear separation of scales no longer exists due to a variable sub-process collision energy. In particular, if the energy is close to or higher than the suppression scale of the operators, we cannot describe the process by a truncated effective operator series anymore.

We propose a programme to address this problem, which is based on a specific relation between the EFT operators and the corresponding simplified models which match onto them. The basic idea is to consider a tree-level process mediated by a new heavy particle whose mass is fixed to be the suppression scale $\Lambda_a$, where the operator ${\cal O}_a$ is generated in the large $\Lambda_a$ limit. This can be realized in a simplified model setting, with all the vertices being now from mass dimension-three or four terms. For each operator ${\cal O}_a$ in Tab.\,\ref{tab: operators}, there are two different ways to introduce such a new heavy particle; in one case the new particle is even under the $Z_2$ symmetry, and in the other case it's $Z_2$-odd. The difference between the two corresponds to which process reproduces ${\cal O}_a$, an $s$-channel or $t$-channel process. In what follows, we consider each of these ways in further detail for each operator.

\subsection{The four-Fermi operators ${\cal O}_f$}
\label{subsec: 4-Fermi}

We first consider the four-Fermi operators ${\cal O}_f = (\bar{\chi} \gamma^\mu \gamma_5 \chi)(\bar{f}_i \gamma_\mu f_j)$ with $f_i$ being SM fermions $Q_i$, $U_i$, $D_i$, $L_i$ and $E_i$, as this is a simple case to intuitively understand the relation between the EFT and simplified models. 

\subsubsection{$Z_2$-even mediator}

The four-Fermi operator $(\bar{\chi} \gamma^\mu \gamma_5 \chi)(\bar{f}_i \gamma_\mu f_j)$ describes, for example, the process $\chi\chi \leftrightarrow f\bar{f}$, so that the new heavy $Z_2$-even particle must be bosonic, being exchanged in the $s$-channel. Since the four-Fermi operator has the form of a current-current interaction, this particle should also be a massive vector. Furthermore, the vector particle directly couples to the dark matter current, and thus it must be a singlet under the SM gauge symmetry. We therefore introduce a real singlet Proca field $(X_f)_\mu$ whose mass is $\Lambda_f$:
\begin{eqnarray}
{\cal L}_f^{(+)} = -\frac{1}{4} (X_f)_{\mu\nu} (X_f)^{\mu\nu} + \frac{\Lambda_f^2}{2} (X_f)_\mu (X_f)^\mu
-\frac{d^{(X)}_\chi}{2} (\bar{\chi} \gamma^\mu \gamma_5 \chi) (X_f)_\mu
-d^{(X)}_f (\bar{f} \gamma^\mu f) (X_f)_\mu,
\label{eq: SM Four-Fermi +}
\end{eqnarray}
where the superscript $(+)$ indicates that $(X_f)_\mu$ is even under the $Z_2$ symmetry, while the index $f$ corresponds to $Q$, $U$, $D$, $L$ or $E$. The flavour index for the SM fermions has been suppressed for simplicity. The field strength tensor of the Proca field is denoted by $(X_f)_{\mu\nu}$, and at this stage, the coupling constants $d^{(X)}_\chi$ and $d^{(X)}_f$ can take arbitrary values, which will be fixed by the relation between the EFT and this simplified model.

After integrating $(X_f)_\mu$ out from the simplified model Lagrangian (\ref{eq: SM Four-Fermi +}), we obtain the following effective Lagrangian which involves a non-local interaction term,\footnote{We performed the integration explicitly in the path-integral form. See Ref.\,\cite{Haba:2011vi} for more details.}
\begin{eqnarray}
	{\cal L}_{f, {\rm eff}}^{(+)} &=&
	\frac{i}{2} \int d^4y\,J_{X_f}^\mu(x)\,{\cal G}_{\mu\nu}^{(X_f)}(x - y)\,J_{X_f}^\nu(y),
	\nonumber \\
	J_{X_f}^\mu(x) &=& (d^{(X)}_\chi/2)
	[\bar{\chi}(x) \gamma^\mu\gamma_5 \chi(x)] + d^{(X)}_f[\bar{f}(x)\gamma^\mu f(x)],
\label{eq: SM Four-Fermi + eff}
\end{eqnarray}
where ${\cal G}_{\mu\nu}^{(X_f)}(x - y)$ is the Green's function (the two-point function) of $(X_f)_\mu$ in the coordinate space and has the following asymptotic form in the limit of large $\Lambda_f$,
\begin{eqnarray}
	{\cal G}_{\mu\nu}^{(X_f)}(x - y) \equiv -i\int \frac{d^4q}{(2\pi)^4}
	\frac{g_{\mu\nu} - q_\mu q_\nu/\Lambda_f^2}{q^2 - \Lambda_f^2 + i\epsilon} e^{-i q(x - y)}
	\to i \frac{g_{\mu\nu}}{\Lambda_f^2}\,\delta(x - y).
\end{eqnarray}
The Lagrangian\,(\ref{eq: SM Four-Fermi + eff}) generates the four-Fermi operator $(\bar{\chi} \gamma^\mu \gamma_5 \chi)(\bar{f} \gamma_\mu f)$ with the coefficient $-d^{(X)}_\chi d^{(X)}_f/(2\Lambda_f^2)$. Since we are interested in tree-level diagrams involving both the DM and the SM particles, physical quantities depend only on the combination of $d^{(X)}_\chi d^{(X)}_f$. Hence, each coupling constant can be fixed without loss of generality as
\begin{eqnarray}
	d^{(X)}_\chi = -1 \qquad {\rm and} \qquad d^{(X)}_f = c_f.
\end{eqnarray}
This defines the required relation between the EFT and the simplified model\,(\ref{eq: SM Four-Fermi +}).

\subsubsection{$Z_2$-odd mediator}

The process $\chi\chi \leftrightarrow f\bar{f}$ is reproduced by the $t(u)$-channel exchanges of a new heavy $Z_2$-odd particle, and the new particle must to be bosonic. On the other hand, unlike the previous case, this new $Z_2$-odd bosonic particle can be either a massive scalar or a vector because of the nature of $t(u)$-channel diagrams. Moreover, since the new bosonic field couples to one DM field and one SM fermion field at each vertex, its quantum numbers must be the same as those of the SM fermion. We therefore introduce a complex scalar field ($\phi_f$) and a complex Proca field $(V_f)_\mu$ simultaneously, where their masses and the quantum numbers are fixed to be $\Lambda_f$ and those of $f$, respectively:
\begin{eqnarray}
	{\cal L}_f^{(-)} &=&
	-\frac{1}{2} (V_f)^\dagger_{\mu\nu} (V_f)^{\mu\nu} + \Lambda_f^2 (V_f)^\dagger_\mu (V_f)^\mu
	-d^{(V)}_f \bar{f} \gamma^\mu \chi (V_f)_\mu
	-(d^{(V)}_f)^* (V_f)^\dagger_\mu \bar{\chi} \gamma^\mu f
	\nonumber \\
	&&
	+|D_\mu \phi_f|^2 - \Lambda_f^2 |\phi_f|^2
	-d^{(\phi)}_f \bar{f} \chi \phi_f
	-(d^{(\phi)}_f)^* \phi_f^\dagger \bar{\chi} f.
\label{eq: SM Four-Fermi -}
\end{eqnarray}
The SM gauge indices for $f$, $(V_f)_\mu$ and $\phi_f$ as well as the flavour index for $f$ have been suppressed for the sake of simplicity. The field strength tensor of the new vector field is given by $(V_f)_{\mu\nu} \equiv D_\mu (V_f)_\nu - D_\nu (V_f)_\mu$ with $D_\mu$ being the covariant derivative which is common for all of the three fields $f$, $(V_f)_\mu$ and $\phi_f$.

After integrating the heavy vector $(V_f)_\mu$ and scaler $\phi_f$ fields out from the above simplified model Lagrangian (\ref{eq: SM Four-Fermi -}), we obtain the following effective Lagrangian,
\begin{eqnarray}
	&&{\cal L}_{f, {\rm eff}}^{(-)} =
	i \int d^4y \left[J_{V_f}^{\mu\dagger}(x)\,{\cal G}_{\mu\nu}^{(V_f)}(x - y)\,J_{V_f}^\nu(y)
	+J_{\phi_f}^\dagger (x)\,{\cal G}^{(\phi_f)}(x - y)\,J_{\phi_f}(y) \right],
	\nonumber \\
	&& J_{V_f}^\mu(x) =
	(d^{(V)}_f)^* \bar{\chi}(x) \gamma_\mu f(x),
	\qquad
	J_{\phi_f}(x) = (d^{(\phi)}_f)^* \bar{\chi}(x) f(x),
	\label{eq: SM Four-Fermi - eff}
\end{eqnarray}
where, as before, ${\cal G}_{\mu\nu}^{(V_f)}(x - y)$ and ${\cal G}^{(\phi_f)}(x - y)$ are the Green's functions of $(V_f)_\mu$ and $\phi_f$, respectively. Here, we have neglected the gauge interaction terms of the massive fields to derive the above effective Lagrangian, because they contribute only to operators of mass dimension higher than six. Taking the large $\Lambda_f$ limit of the Green's functions in Eq.\,(\ref{eq: SM Four-Fermi - eff}), we obtain the following four-Fermi operators:
\begin{eqnarray}
	{\cal L}_{f, {\rm eff}}^{(-)} \to \pm\frac{|d^{(\phi)}_f|^2 - 2 |d^{(V)}_f|^2}{4\Lambda_f^2}
	(\bar{\chi}\gamma^\mu\gamma_5\chi)(\bar{f}\gamma_\mu f),
\end{eqnarray}
where Fierz transformations have been used to derive the limit.\footnote{Other types of four-Fermi operators generated by the Fierz transformations such as $\bar{\chi} \gamma^\mu \chi$ vanish identically because of the Majorana nature of the DM field and the chiral nature of the SM fermion fields.} Regarding the sign of the equation, the `$+$' sign should be used for $f = U$, $D$ and $E$, while the `$-$' sign should be used for $f = Q$ and $L$. The introduction of both the massive Proca and scalar fields simultaneously is necessary to make the coefficient of the operator $(\bar{\chi} \gamma^\mu \gamma_5 \chi)(\bar{f} \gamma_\mu f)$ take both positive and negative values. Both the coupling constants $d^{(V)}_f$ and $d^{(\phi)}_f$ can be real, for their phases can be removed by the redefinitions of the fields $(V_f)_\mu$ and $\phi_f$. Thus the relation between the EFT and the simplified model (\ref{eq: SM Four-Fermi -}) can be expressed as
\begin{eqnarray}
	&& d^{(V)}_f = \sqrt{-c_f}\,\theta(-c_f)
	\quad {\rm and} \quad
	d^{(\phi)}_f = \sqrt{2c_f}\,\theta(c_f)
	\qquad {\rm for}\quad
	f = U, D~{\rm and}~E,
	\nonumber \\
	&& d^{(V)}_f = \sqrt{c_f}\,\theta(c_f)
	\quad {\rm and} \quad
	d^{(\phi)}_f = \sqrt{-2c_f}\,\theta(-c_f)
	\qquad {\rm for} \quad f = Q~{\rm and}~L.
\end{eqnarray}

\subsection{The operator ${\cal O}_H$}
\label{subsec: Operator H}

We next consider the operator, ${\cal O}_H = (\bar{\chi} \gamma^\mu \gamma_5 \chi)(H^\dagger i\overleftrightarrow{D_\mu} H)$, involving the DM and the Higgs fields, which plays an important role in DM signals at colliders. 

\subsubsection{$Z_2$-even mediator}

The operator describes, e.g., the process $\chi\chi \leftrightarrow H H^\dagger$, so that the new heavy $Z_2$-even particle must be a boson connecting the DM current $\bar{\chi} \gamma^\mu \gamma_5 \chi$ and the Higgs boson current $H^\dagger i\overleftrightarrow{D_\mu} H = iH^\dagger D_\mu H - i(D_\mu H)^\dagger H$, which is almost the same situation as the previous case. Hence, we introduce a real singlet Proca field $(X_H)_\mu$ with the mass of $\Lambda_H$:
\begin{eqnarray}
	{\cal L}_H^{(+)} &=& -\frac{1}{4} (X_H)_{\mu\nu} (X_H)^{\mu\nu}
	+\frac{\Lambda_H^2}{2} (X_H)_\mu (X_H)^\mu
	\nonumber \\
	&&
	-\frac{d^{(X)}_\chi}{2} (\bar{\chi} \gamma^\mu \gamma_5 \chi) (X_H)_\mu
	+|\{D_\mu + i d^{(X)}_H (X_H)_\mu\}H|^2,
	\label{eq: SM H +}
\end{eqnarray}
where $D_\mu$ is the SM covariant derivative acting on the Higgs field $H$. The kinetic term of $H$ gives two interactions, $-d^{(X)}_H (H^\dagger i\overleftrightarrow{D_\mu} H) (X_H)^\mu$ and $(d^{(X)}_H)^2 (X_H)_\mu (X_H)^\mu |H|^2$. Because the latter term contributes only to operators of mass dimension more than six, we shall drop it. After integrating the Proca field $(X_H)_\mu$ out from the above simplified model Lagrangian (\ref{eq: SM H +}), we obtain the following effective Lagrangian,
\begin{eqnarray}
	{\cal L}_{H, {\rm eff}}^{(+)} &=&
	\frac{i}{2} \int d^4y\,J_{X_H}^\mu(x)\,{\cal G}_{\mu\nu}^{(X_H)}(x - y)\,J_{X_H}^\nu(y),
	\nonumber \\
	J_{X_H}^\mu(x) &=&
	(d^{(X)}_\chi/2)[\bar{\chi}(x) \gamma^\mu\gamma_5 \chi(x)]
	+d^{(X)}_H[H^\dagger(x) i\overleftrightarrow{D_\mu} H(x)],
	\label{eq: SM H + eff}
\end{eqnarray}
Taking a large $\Lambda_H$ limit of the Green's function, the operator $(\bar{\chi} \gamma^\mu \gamma_5 \chi)(H^\dagger i\overleftrightarrow{D_\mu} H)$ is obtained with its coefficient being $-d^{(X)}_\chi d^{(X)}_H/(2\Lambda_H^2)$. With the same reasoning as before, the relation between the EFT and the simplified model\,(\ref{eq: SM H +}) can be written as
\begin{eqnarray}
d^{(X)}_\chi = -1 \qquad {\rm and} \qquad d^{(X)}_H = c_H.
\end{eqnarray}

\subsubsection{$Z_2$-odd mediator}

The process $\chi\chi \leftrightarrow H H^\dagger$ can take place via the $t(u)$-channel exchange of a new heavy $Z_2$-odd particle. Hence, the new particle must be a fermion having the same quantum numbers as those of the Higgs boson in this case. We therefore introduce a Dirac fermion field (denoted by $\psi_H$) with mass $\Lambda_H$ in the following simplified model:
\begin{eqnarray}
	{\cal L}_H^{(-)} =
	\bar{\psi}_H (i\slashed{D} - \Lambda_H)\psi_H
	-\bar{\chi} H^\dagger [d^{(\psi)}_{H1} + id^{(\psi)}_{H2}\gamma_5] \psi_H
	-\bar{\psi}_H [(d^{(\psi)}_{H1})^* + i(d^{(\psi)}_{H2})^* \gamma_5] H \chi,
	\label{eq: SM H -}
\end{eqnarray}
where the covariant derivative $D_\mu$ acting on the new fermion field $\psi_H$ is exactly the same as the one acting on the Higgs field $H$. After integrating $\psi_H$ out from the above simplified model Lagrangian (\ref{eq: SM H -}), we obtain the following effective Lagrangian,
\begin{eqnarray}
	&& {\cal L}_{H, {\rm eff}}^{(-)} =
	i \int d^4y\,\bar{J}_{\psi_H}(x)\,{\cal G}^{(\psi_H)}(x - y)\,J_{\psi_H}(y)
	\label{eq: SM H - eff} \\
	&&\qquad~~~~
	+\int d^4y\,d^4z\,\bar{J}_{\psi_H}(x)\,{\cal G}^{(\psi_H)}(x - y)
	\left[ \frac{g}{2} \slashed{W}^a(y) \sigma^a + \frac{g'}{2} \slashed{B}(y) \right]
	{\cal G}^{(\psi_H)}(y - z)\,J_{\psi_H}(z),
	\nonumber \\
	&& J_{\psi_H}(x) =
	[(d^{(\psi)}_{H1})^* + i(d^{(\psi)}_{H2})^* \gamma_5] H(x) \chi(x),
	\qquad
	\bar{J}_{\psi_H}(x) =\bar{\chi}(x) H^\dagger(x) [d^{(\psi)}_{H1} + id^{(\psi)}_{H2}\gamma_5],
	\nonumber
\end{eqnarray}
where $\slashed{W}^a$ and $\slashed{B}$ are the SU(2)$_L$ and U(1)$_Y$ gauge fields, respectively, with $g$ and $g'$ being their gauge couplings. Here again we have neglected other terms which contribute only to operators of mass dimension more than six. Then, the Green's function has the following asymptotic form in the limit of large $\Lambda_H$,
\begin{eqnarray}
	{\cal G}^{(\psi_H)}(x - y) \equiv i\int \frac{d^4q}{(2\pi)^4}
	\frac{\slashed{q} + \Lambda_H}{q^2 - \Lambda_H^2 + i\epsilon} e^{-i q(x - y)}
	\to -\frac{i}{\Lambda_H}\,\delta(x - y) + \frac{1}{\Lambda_H^2} \slashed{\partial}_x \delta(x - y).
\end{eqnarray}

As a result, the effective Lagrangian\,(\ref{eq: SM H - eff}) has the following asymptotic form in the limit, which involves not only the operator $(\bar{\chi} \gamma^\mu \gamma_5 \chi)(H^\dagger i\overleftrightarrow{D_\mu} H)$, but also other ones:
\begin{eqnarray}
	&& {\cal L}_{H, {\rm eff}}^{(-)} \to
	\frac{1}{\Lambda_H} \Im[(d^{(\psi)}_{H1})^*\,d^{(\psi)}_{H2}]
	(\bar{\chi} \gamma^\mu \gamma_5 \chi)(H^\dagger i\overleftrightarrow{D_\mu} H)
	+\frac{2}{\Lambda_H}\Re[(d^{(\psi)}_{H1})^*\,d^{(\psi)}_{H2}]
	(\bar{\chi}i\gamma_5\chi) |H|^2
	\nonumber \\
	&& \qquad~~~
	+\frac{1}{\Lambda_H^2}
	[(\Lambda_H + M_\chi)|d^{(\psi)}_{H1}|^2 - (\Lambda_H - M_\chi)|d^{(\psi)}_{H2}|^2]
	(\bar{\chi}\chi) |H|^2.
	\label{eq: SM H - eff limit}
\end{eqnarray}
where the equation of motion for the DM field, $i\slashed{\partial} \chi = M_\chi\chi + \cdots$, has been used to obtain the above result. One of the phases of the coupling constants $d^{(\psi)}_{H1}$ and $d^{(\psi)}_{H2}$ can be removed by the redefinition of the Dirac field $\psi_H$. By imposing the condition that the coefficients of the other operators $(\bar{\chi}i\gamma_5\chi) |H|^2$ and $(\bar{\chi}\chi) |H|^2$ are zero, the relation between the EFT and the simplified model (\ref{eq: SM H -}) is obtained as
\begin{eqnarray}
	d^{(\psi)}_{H1} =
	\frac{1}{\sqrt{2}}\left(\frac{\Lambda_H - M_\chi}{\Lambda_H + M_\chi}\right)^{1/4} |c_H|^{1/2}
	\qquad {\rm and} \qquad
	d^{(\psi)}_{H2} =
	\frac{i}{\sqrt{2}}\left(\frac{\Lambda_H + M_\chi}{\Lambda_H - M_\chi}\right)^{1/4}
	\frac{c_H}{|c_H|^{1/2}}.
\end{eqnarray}

The emergence of the operators in Eq.\,(\ref{eq: SM H - eff limit}) can be interpreted as a consequence of the mixing between $\chi$ and $\psi_H$. In fact, $(\bar{\chi} \gamma^\mu \gamma_5 \chi)(H^\dagger i\overleftrightarrow{D_\mu} H)$ describes a vertex between the $Z$ boson and two $\chi$ fields after the electroweak symmetry breaking (EWSB), and thus the gauge interaction of $\chi$. The DM particle is no longer a pure singlet under the SM gauge symmetry, but acquires a small doublet component after the EWSB.

\subsection{The scalar interaction operator ${\cal O}_S$}
\label{subsec: Operator S}

Here, we consider the dimension-five operator ${\cal O}_H = (\bar{\chi} \chi) |H|^2$, which is relevant, in the context of colliders, to the decay of the Higgs boson into a pair of DM particles.

\subsubsection{$Z_2$-even mediator}

This operator describes the process $\chi\chi \leftrightarrow H H^\dagger$ again, and thus the new heavy $Z_2$-even particle must be bosonic. Since it connects the DM and the Higgs scalar operators instead of their currents, we introduce a real singlet scalar field $\varphi_S$ with mass $\Lambda_S$:
\begin{eqnarray}
	{\cal L}_S^{(+)} =
	\frac{1}{2} (\partial_\mu \varphi_S)^2 - \frac{\Lambda_S^2}{2} \varphi_S^2
	-\frac{d^{(\varphi)}_\chi}{2} \varphi_S (\bar{\chi}\chi)
	-d^{(\varphi)}_S \Lambda_S \varphi_S |H|^2.
	\label{eq: SM S +}
\end{eqnarray}
Other renormalizable terms in the Lagrangian, which are not relevant in deriving ${\cal O}_S$, are not shown here. After integrating the scalar field $\varphi_S$ out from the above simplified model Lagrangian (\ref{eq: SM S +}), we obtain the following effective Lagrangian:
\begin{eqnarray}
	&& {\cal L}_{S, {\rm eff}}^{(+)} =
	\frac{i}{2} \int d^4y\,J_{\varphi_S}(x)\,{\cal G}^{(\varphi_S)}(x - y)\,J_{\varphi_S}(y),
	\nonumber \\
	&& J_{\varphi_S}(x) =
	(d^{(\varphi)}_\chi/2)[\bar{\chi}(x)\chi(x)] + d^{(\varphi)}_S \Lambda_S |H(x)|^2.
	\label{eq: SM S + eff}
\end{eqnarray}
Taking the large $\Lambda_S$ limit, the effective Lagrangian\,(\ref{eq: SM S + eff}) generates the operator $(\bar{\chi}\chi)|H|^2$ with its coefficient being $d^{(\varphi)}_\chi d^{(\varphi)}_S/(2\Lambda_S)$. Thus, the relation between the EFT and the simplified model\,(\ref{eq: SM S +}) can be obtained as:
\begin{eqnarray}
	d^{(\varphi)}_\chi = 1 \qquad {\rm and} \qquad d^{(\varphi)}_S = c_S.
\end{eqnarray}

\subsubsection{$Z_2$-odd mediator}

The process $\chi\chi \leftrightarrow H H^\dagger$ is generated by the $t(u)$-channel exchange of a new heavy $Z_2$-odd particle, and the new particle must be a fermion having the same quantum numbers as those of $H$. We therefore introduce a Dirac fermion field ($\psi_S$) with mass $\Lambda_S$:
\begin{eqnarray}
	{\cal L}_S^{(-)} =
	\bar{\psi}_S (i\slashed{D} - \Lambda_S)\psi_S
	-\bar{\chi} H^\dagger [d^{(\psi)}_{S1} + id^{(\psi)}_{S2}\gamma_5] \psi_S
	-\bar{\psi}_S [(d^{(\psi)}_{S1})^* + i(d^{(\psi)}_{S2})^* \gamma_5] H \chi,
	\label{eq: SM S -}
\end{eqnarray}
where $D_\mu$ is the covariant derivative acting on the new fermion field $\psi_S$. Since the simplified model (\ref{eq: SM S -}) is exactly the same as the one in Eq.\,(\ref{eq: SM H -}), it gives rise to the three higher dimensional operators $(\bar{\chi}\chi) |H|^2$, $(\bar{\chi}i\gamma_5\chi) |H|^2$ and $(\bar{\chi} \gamma^\mu \gamma_5 \chi)(H^\dagger i\overleftrightarrow{D_\mu} H)$ after integrating the heavy field $\psi_S$ out from the simplified model Lagrangian (\ref{eq: SM S -}) and taking a large $\Lambda_S$ limit. As a result, by imposing the coefficients of the latter two operators to be zero, the relation between the EFT and the simplified model (\ref{eq: SM S -}) is obtained as
\begin{eqnarray}
	d^{(\psi)}_{S1} =
	\frac{1}{\sqrt{2}}\left(\frac{\Lambda_S\,c_S}{\Lambda_S + M_\chi}\right)^{1/2} \theta(c_S)
	\qquad {\rm and} \qquad
	d^{(\psi)}_{S2} =
	\frac{1}{\sqrt{2}}\left(\frac{-\Lambda_S\,c_S}{\Lambda_S - M_\chi}\right)^{1/2} \theta(-c_S).
\end{eqnarray}

\subsection{The pseudoscalar interaction operator ${\cal O}_{PS}$}
\label{subsec: Operator PS}

Finally, we consider the other (CP violating) dimension-five operator ${\cal O}_{PS} = (\bar{\chi} i \gamma_5 \chi) |H|^2$. 

\subsubsection{$Z_2$-even mediator}

It is essentially the same as in the case of the scalar operator ${\cal O}_S$ in the previous subsection. We therefore introduce a real singlet scalar field $\varphi_{PS}$ whose mass is fixed to be $\Lambda_{PS}$, and use the pseudoscalar interaction $\varphi\,(\bar{\chi} i\gamma_5 \chi)$ instead of the scalar one $\varphi\,(\bar{\chi} \chi)$:
\begin{eqnarray}
	{\cal L}_{PS}^{(+)} =
	\frac{1}{2} (\partial_\mu \varphi_{PS})^2
	-\frac{\Lambda_{PS}^2}{2} \varphi_{PS}^2
	-\frac{d^{(\varphi)}_\chi}{2} \varphi_{PS} (\bar{\chi} i\gamma_5 \chi)
	-d^{(\varphi)}_{PS} \Lambda_{PS} \varphi_{PS} |H|^2.
	\label{eq: SM PS +}
\end{eqnarray}
After integrating $\varphi_{PS}$ out from the Lagrangian above and taking a large $\Lambda_{PS}$ limit, we obtain the dimension-five operator $(\bar{\chi} i\gamma_5 \chi)|H|^2$ with coefficient $d^{(\varphi)}_\chi d^{(\varphi)}_{PS}/(2\Lambda_{PS})$. Thus, the relation between the EFT and the simplified model\,(\ref{eq: SM PS +}) is obtained as
\begin{eqnarray}
	d^{(\varphi)}_\chi = 1 \qquad {\rm and} \qquad d^{(\varphi)}_{PS} = c_{PS}.
\end{eqnarray}

\subsubsection{$Z_2$-odd mediator}

This is again similar to the case of the scalar operator ${\cal O}_S$, and we therefore introduce a Dirac fermion field ($\psi_{PS}$) again with its mass being $\Lambda_{PS}$:
\begin{eqnarray}
	{\cal L}_{PS}^{(-)} =
	\bar{\psi}_{PS} (i\slashed{D} - \Lambda_{PS})\psi_{PS}
	-\bar{\chi} H^\dagger [d^{(\psi)}_{PS1} + id^{(\psi)}_{PS2}\gamma_5] \psi_{PS}
	-\bar{\psi}_{PS} [(d^{(\psi)}_{PS1})^* + i(d^{(\psi)}_{PS2})^* \gamma_5] H \chi.
	\label{eq: SM PS -}
\end{eqnarray}
It generates the three higher dimensional operators in exactly the same manner as before. By imposing the coefficients of the operators $(\bar{\chi} \gamma^\mu \gamma_5 \chi)(H^\dagger i\overleftrightarrow{D_\mu} H)$ and $(\bar{\chi}\chi) |H|^2$ to be zero, the relation between the EFT and this simplified model is obtained as
\begin{eqnarray}
	d^{(\psi)}_{PS1} =
	\left(\frac{\Lambda_{PS} - M_\chi}{\Lambda_{PS} + M_\chi}\right)^{1/4} \frac{|c_{PS}|^{1/2}}{2}
	\qquad {\rm and} \qquad
	d^{(\psi)}_{PS2} =
	\left(\frac{\Lambda_{PS} + M_\chi}{\Lambda_{PS} - M_\chi}\right)^{1/4}
	\frac{c_{PS}}{2|c_{PS}|^{1/2}}.
\end{eqnarray}

\subsection{Some caveats to using simplified models}
\label{subsec: caveat}

In UV completions of models with vector mediators, one needs to consider the questions of gauge invariance, the origin of the vector boson mass, perturbative unitarity of scattering amplitudes and cancellation of anomalies. For an s-channel UV completion with a new $U(1)^\prime$ gauge group (and a massive $Z^\prime$ boson), a detailed analysis has been carried out in Ref.\,\cite{Kahlhoefer:2015bea}. However, with our EFT assumption of $m_\chi < m_{Z^\prime}$, the process leading to strong unitarity constraint $\chi \chi \to Z_L^\prime Z_L^\prime$ is kinematically forbidden, where $Z_L^\prime$ refers to the longitudinally polarized state of $Z^\prime$ . From unitarity of $\chi \chi \to f \bar{f}$ and similar 2-to-2 interactions between pairs of SM fermions ($f$) and $\chi$, one again finds the constraint $m_\chi, m_f < m_{Z^\prime}$, both of which are satisfied for the region of our interest $\Lambda > {\rm Max}[3\,m_{\rm DM}$, 0.3\,{\rm TeV}] (see next section for details on the minimum value of $\Lambda$). The only constraint that applies therefore is that of gauge invariance under the new $U(1)^\prime$, when left and right handed SM fermions have different charges under the new gauge group. In this case, the SM Higgs doublet needs to be charged under $U(1)^\prime$ as well, implying that if the SM quarks are charged, so are the SM leptons, leading to stronger constraints from LHC dilepton resonance searches. Such a conclusion can however be avoided if the SM charged lepton and quark masses are generated by two different Higgs doublets, thereby making their $U(1)^\prime$ charges uncorrelated. In the spirit of this study, we would resort to such an UV completion, in order not to over-constrain the general picture for DM. To summarize, specific UV models can be more constrained by the requirements of gauge invariance or unitarity, however, in the domain of validity of the EFT, unitarity constraints are satisfied, and gauge invariance can be achieved with, for e.g., a non-minimal Higgs sector in order to avoid stronger dilepton constraints. Moreover, as is well known, unless we consider very specific $U(1)^\prime$ charge assignments, there would be additional gauge anomalies in the theory, and to cancel such anomalies we have to introduce new heavy quarks which are vector-like under the SM gauge group, but are chiral under the new $U(1)^\prime$. Such new quarks can be taken to be heavier than $\Lambda$, and their effect on DM phenomenology is not expected to be generically significant. For the $Z_2$-odd vector boson mediator, the UV completion needs more model building, where we either have to embed the DM particle and the SM fermions in a multiplet of a non-abelian gauge group, or look for an extra-dimensional or Little Higgs type model with an exact KK or T-parity. In the latter two cases, the $Z_2$-odd partner of a right-handed neutrino, for example, can be the fermionic DM candidate. 

The simplified models with scalar mediators involve only new Yukawa-type interactions, and they therefore can be considered UV complete. It should however be ensured that the complete scalar potential, which is not relevant to this study, does not develop any charge or colour breaking minimum. Since we did not consider any mixing between the $SU(2)_L$ singlet and doublet scalars (which would induce dimension 7 operators), charge breaking minimum is not expected to develop.

Finally, certain SM observables are also sensitive to the UV completion of the DM simplified models we consider, one example being the production cross-section and partial decay widths to SM states of the Higgs boson. Even if the mediators introduced are sufficiently heavy compared to the electroweak scale, they can modify these quantities and only a global fit of all current LHC Higgs data can determine whether such deviations are permissible at present. This is not a generic effect though, as for example, the $Z_2$-even mediator for the scalar interaction operator in Sec.\,\ref{subsec: Operator S} can induce such modifications via the mixing of the singlet scalar with the neutral CP-even Higgs field after the electroweak symmetry breaking, but the $Z_2$-odd mediator mixing with the singlet DM state will not have any such effect at the leading order. Even for the $Z_2$-even mediator case, a UV completion can be designed to cancel such modifications predicted by the minimal DM simplified models. Therefore once again, in order not to over-constrain the generic DM picture, we assume the SM predictions for such processes are unaltered in the complete UV theory. The same approach also applies to the four-fermion SM operators generated in the simplified models for DM-SM fermion interactions (at leading order, in the s-channel models), which can furnish strong constraints especially for the first generation quark and charged lepton couplings of the DM, in the absence of accidental cancellations or symmetry protection.

\section{Application to singlet-like Majorana fermion DM}
\label{sec: application}

We are in a position to apply the method in the previous section to a singlet-like Majorana fermion DM, in order to determine the viable regions, in the space spanned by the DM mass $m_{\rm DM}$, the suppressions scales $\Lambda_i$ and the coupling constants $c_i$. Even in the absence of flavour changing interactions of the form $c_{f_{ij}}(\bar{\chi} \gamma^\mu \gamma_5 \chi) (\bar{f}_i \gamma_\mu f_j)$ with $i \neq j$, we have 37 parameters (19 parameters if we only use the EFT\,\cite{Bhattacherjee:2012ch}) to deal with, which is beyond the scope of our numerical analysis. Hence, we need to impose the following simplifying assumptions before attempting a scan of the entire parameter space:
\begin{itemize}
	\setlength{\itemsep}{0mm}

	\item {\boldmath $\Lambda_i = \Lambda$} (unless otherwise stated):
	All the intermediate heavy particles introduced in the simplified models are assumed to have a common mass $\Lambda$.

	\item {\boldmath $\Lambda > {\rm Max}[3m_{\rm DM}, 0.3\,{\rm TeV}]$}:
	In order to describe DM pair annihilation by the EFT, we must have $\Lambda>2\,m_{\rm DM}$ (note that $\Lambda=2\,m_{\rm DM}$ induces $100\%$ error in the theoretical prediction based only on dimension-6 operators for s-channel UV completions). Furthermore, to reduce the error in the prediction of the invisible decays of the $Z$ or the Higgs boson using the leading effective operators, we need $\Lambda > v$, where $v(=246\,{\rm GeV})$ is the scale of electroweak symmetry breaking. Therefore, we choose the minimum value of $\Lambda$ being consistent with these two requirements.

	\item {\boldmath $c_{f_{ij}} = c_f$}:
	The four-Fermi operators have a flavour-blind structure, i.e., $c_f$ does not depend on the flavour indices. On the other hand, SM fermions belonging to different representations are allowed to have different couplings with the DM.

	\item {\boldmath $|c_i| \leq 1$}:
	This reflects the implicit assumption followed throughout this paper: the UV physics behind the EFT is described by a weakly coupled theory.

	\item {\boldmath $c_{PS} = 0$}: CP symmetry is assumed to be preserved in the interactions of the DM particle with the SM particles.

\end{itemize}
Under these conditions, the number of free parameters reduces to 9 ($m_{\rm DM}$, $\Lambda$, $c_S$, $c_H$, $c_Q$, $c_U$, $c_D$, $c_L$ and $c_E$), which is within the scope of our analysis. Even though this is a limited scan of the most general parameter space, it includes sufficient degrees of freedom to grasp the broad picture of a singlet-like Majorana fermion DM scenario. We should mention that if the CP violating pseudoscalar coupling $c_{PS}$ is non-zero, the phenomenology changes considerably as observed in our previous study\,\cite{Matsumoto:2014rxa}, and in contrast to the CP-conserving case, indirect detection experiments become more relevant. 

In order to explore the high probability density regions of the multi-dimensional parameter space, we employ the profile-likelihood method\,\cite{Rolke:2004mj}. All the relevant experimental and observational constraints are incorporated in the likelihood function $L$ taking into account their statistical and systematic uncertainties. The likelihood function that we adopt in our analysis is composed of three parts:
\begin{eqnarray}
	L[m_{\rm DM}, \Lambda, c_S, \cdots, c_E] =
	L_{\rm Cos}[m_{\rm DM}, \Lambda, \cdots] \times
	L_{\rm Det}[m_{\rm DM}, \Lambda, \cdots] \times
	L_{\rm Coll}[m_{\rm DM}, \Lambda, \cdots],
\end{eqnarray}
where these components involve information obtained from DM cosmology, DM detection experiments and collider experiments, respectively. In the following subsections, we shall describe each component in detail. In particular, we carefully discuss how the relation between the EFT and the simplified models is applied to calculate the component $L_{\rm Coll}$. We adopt the MultiNest sampling algorithm\,\cite{Feroz:2008xx}, which is an efficient implementation of the Markov Chain Monte Carlo algorithm\,\cite{Press:1992zz}. While describing our results in the relevant set of two-dimensional parameter space, e.g., the $(m_{\rm DM}, \Lambda)$-space, we maximize the likelihood function along the directions of the other parameters. Our scan of the parameters spans the following ranges, 10\,GeV $\leq m_{\rm DM} \leq$ 5\,TeV and $\Lambda \leq$ 100\,TeV, which are determined based on our previous study\,\cite{Matsumoto:2014rxa}. We use a flat prior for all the operator coefficients, while both flat and log priors are used for different regions in $m_{\rm DM}$ and $\Lambda$, in order to obtain as large a coverage of the whole parameter space as possible.

\subsection{Constraints from DM cosmology}

We adopt the following Gaussian likelihood function $L_{\rm Cos}$ in our analysis:
\begin{eqnarray}
	L_{\rm Cos}[m_{\rm DM}, \Lambda, c_S, \cdots, c_E] \propto
	\theta(\Omega_{\rm OBS} - \Omega_{\rm TH})
	+\exp\left[-\frac{(\Omega_{\rm TH} - \Omega_{\rm OBS})^2}{2\,(\delta\Omega)^2}\right]
	\theta(\Omega_{\rm TH} - \Omega_{\rm OBS}),
	\label{eq: LCos}
\end{eqnarray}
where $\Omega_{\rm OBS} = 0.1198/h^2$ (with $h$ being the normalized Hubble constant) is the observed cosmological DM parameter obtained in the latest PLANCK results\,\cite{Ade:2015xua}, while $\delta \Omega$ is the error including both the observational and theoretical uncertainties as $(\delta\Omega)^2 = (0.0015/h^2)^2 + (0.025\,\Omega_{\rm TH})^2$. The theoretical uncertainty of $2.5\%$ in the computation of $\Omega_{\rm TH}$ originates from the temperature dependence of massless degrees of freedom required to solve the Boltzmann equation\,\cite{Drees:2015exa}. In order to evaluate $\Omega_{\rm TH}$, we assume the initial condition of thermal equilibrium abundance of the DM particles and compute $\Omega_{\rm TH}$ using the {\tt micrOMEGAs} code\,\cite{Belanger:2006is, Belanger:2008sj, Belanger:2010gh, Belanger:2013oya} with the input model files for {\tt CalcHEP}\,\cite{Belyaev:2012qa} generated by {\tt FeynRules}\,\cite{Christensen:2008py, Alloul:2013bka}. Since all interactions of the EFT in Eq.\,(\ref{eq: EFT}) contribute to $\Omega_{\rm TH}$, the likelihood function $L_{\rm Cos}$ is a function of all the 9 parameters, $m_{\rm DM}$, $\Lambda$, $c_S$, $\cdots$, $c_E$.

As we can see from Eq.\,(\ref{eq: LCos}), the observed value of the relic abundance $\Omega_{\rm OBS}$ is taken only as an upper bound. However, we assume that a single DM species makes up the entire relic abundance. Therefore, if a parameter point gives $\Omega_{\rm TH} < \Omega_{\rm OBS}$, we need an additional non-thermal production giving rise to rest of the required DM density. Though such a non-thermal mechanism is not described by the EFT, it can exist in the UV completion without changing the DM phenomenology discussed here. A typical example is the late time decay of gravitino into neutralino DM in a supersymmetric scenario. 

\subsection{Constraints from DM detection experiments}

We first define the likelihood function $L_{\rm Det}$ using the latest results of direct detection experiments. Even though current indirect detection probes turn out to be much less important compared to other constraints for the singlet-like Majorana fermion DM, we briefly point out the cases where non-negligible constraints might arise in the near future.

\subsubsection{The likelihood function $L_{\rm Det}$}

This likelihood function is further decomposed into three components $L_{\rm Det} = L_{\rm SI} \times L_{\rm SDp} \times L_{\rm SDn}$, where the components correspond to spin-independent scatterings with a nucleon (proton or neutron), and spin-dependent scatterings with a proton and with a neutron, respectively. Since no viable evidence of a DM signal has been obtained yet in direct detection experiments, we adopt the following likelihood function for all the components by taking a Gaussian form with a mean value of zero:
\begin{eqnarray}
	&& L_{\rm Det} \propto
	L_{\rm SI}[m_{\rm DM}, \Lambda, c_S]
	\times L_{\rm SDp}[m_{\rm DM}, \Lambda, c_H, c_Q, c_U, c_D] 
	\times L_{\rm SDn}[m_{\rm DM}, \Lambda, c_H, c_Q, c_U, c_D],
	\nonumber \\
	&&
	L_{\rm SI} = \exp\left[-\frac{\sigma_{\rm SI}^2}{2(\delta\sigma_{\rm SI})^2}\right],
	L_{\rm SDp} = \exp\left[-\frac{\sigma_{\rm SDp}^2}{2(\delta\sigma_{\rm SDp})^2}\right],
	L_{\rm SDn} = \exp\left[-\frac{\sigma_{\rm SDn}^2}{2(\delta\sigma_{\rm SDn})^2}\right],
	\label{eq: LDet}
\end{eqnarray}
where $\sigma_i$ (with $i$ being SI, SDp or SDn) is the corresponding scattering cross-section predicted by the EFT, while $\delta \sigma_i$ is given by the experimental limit on $\sigma_i$ as well as its theoretical uncertainty, which is evaluated as $\delta \sigma_i^2 = {\rm UL}_i^2/1.64^2 + (0.2\,\sigma_i)^2$. Here, ${\rm UL}_i$ stands for the upper limit on $\sigma_i$ at 90\% confidence level, which is given as a function of $m_{\rm DM}$ by an appropriate direct detection experiment. The scaling factor 1.64 is required to make the limit to be the one at 1$\sigma$ level. For the limit on the SI scattering cross-section, we use the result of the LUX experiment (Fig.\,5 in Ref.\,\cite{Akerib:2013tjd}), while the results of the PICO-2L (Fig.\,6 in Ref\,\cite{Amole:2015lsj}) and the XENON100 (Fig.\,2 in Ref.\,\cite{Aprile:2013doa}) experiments are used to put limits on the SDp and SDn scattering cross-sections, respectively.

An additional theoretical uncertainty of $20\%$ is introduced in $\delta \sigma_i$, which originates in hadron matrix elements required to compute $\sigma_i$. Other theoretical uncertainties also come from the local velocity distribution and the local mass density of the DM. Though the former one gives a smaller uncertainty than that of the matrix elements when $m_{\rm DM}$ is large enough, the latter one may not be small\,\cite{Catena:2011kv}. All the experimental results are, however, given assuming the local mass density of 0.3\,GeV/cm$^3$, while recent analysis of the Milky Way mass models show that the density is higher than this value\,\cite{Nesti:2013uwa}. We therefore do not include these astrophysical uncertainties, for the experimental results turn out to always give conservative limits. The scattering cross-section $\sigma_i$ is computed using the {\tt micrOMEGAs} code with its default setting. Here, it is worth pointing out that $L_{\rm SI}$ depends only on $m_{\rm DM}$, $\Lambda$ and $c_S$, because the SI scattering always occurs by exchanging the Higgs boson within the EFT including operators up to mass dimension-six.\footnote{The SI scattering cross-section with a proton is almost the same as the one with a neutron.} On the other hand, all axial current interactions of the DM contribute to the SD scatterings, so that $L_{\rm SDp}$ and $L_{\rm SDn}$ depend on $c_H$, $c_Q$, $c_U$ and $c_D$ in addition to $m_{\rm DM}$ and $\Lambda$.

\subsubsection{Constraints from indirect detection experiments}

We estimate how stringent indirect detection constraints could be for the singlet-like Majorana fermion DM. Since the CP violating interaction is not included in our analysis, all non-relativistic annihilation cross-sections of the DM are suppressed at leading order due to CP and angular momentum conservations (i.e., due to helicity suppression). Though DM indirect detection experiments are usually expected to give less stringent constraints than other detections, they could be important in the following cases:
\begin{itemize}
	\setlength{\itemsep}{0mm}

	\item
	{\bf DM annihilation to a top quark pair} can occur via the s-channel exchange of the $Z$ boson from the interaction ${\cal O}_H$ and also by the four-Fermi interactions ${\cal O}_Q$ and ${\cal O}_U$.\footnote{Amplitude from the s-channel $H$-exchange vanishes in the non-relativistic limit for the DM momenta.} Its annihilation cross-section is estimated to be $\sigma v \sim c_i^2\,m_t^2/\Lambda^4$ (with $c_i$ being the coefficient of the operator ${\cal O}_i$), which is below the current upper limit of DM indirect detection experiments, for values of $\Lambda > 3 m_\chi$.

	 	 \item {\bf DM annihilation to a bottom quark pair} may be constrained if $m_{\rm DM} = {\cal O}(10)$\,GeV. The process is essentially the same as the one to a top quark pair, so that its cross-section is evaluated as $\sigma v \sim c_i^2\,m_b^2/\Lambda^4$ (with $m_b$ being the bottom quark mass). The coefficient $c_i$ is, however, severely constrained by collider experiments when $m_{\rm DM} = {\cal O}(10)$\,GeV as we shall see in the next subsection, and $\sigma v$ is again below the current upper limit, for values of $\Lambda > 300$\,GeV. 
 
	\item {\bf At next to leading order}, several DM annihilation processes do not suffer from helicity suppression, hence they can lead to detectable signals. These include internal bremsstrahlung\,\cite{Bergstrom:1988fp, Bringmann:2007nk} and final state radiation of a photon\,\cite{Bergstrom:1989jr}, from charged intermediate and final states. Though both these processes are suppressed by an additional power of the fine structure constant, they can be enhanced due to collinear singularity. Though the relevant cross-section is currently below the upper limit of DM indirect detection experiments, but the process will be eventually important in the near future as it predicts a hard photon spectrum.
	
	\item {\bf Some interactions of mass dimension more than six} may give DM annihilation signals. Such higher dimensional interactions are usually much suppressed by larger powers of the scale $\Lambda$. The exception would be the operator of $(c/\Lambda^3)(\bar{\chi}\chi)\times$(Yukawa interaction). After the Higgs field acquires its vacuum expectation value, it induces the annihilation $\chi \chi \to f \bar{f}$ (SM fermions) without the helicity suppression. When the coupling $c$ is ${\cal O}(1)$, which requires a large mixing effect between left- and right-handed chiralities, the operator can be responsible for the DM relic abundance and can lead to signals in DM indirect detection experiments.

	\item {\bf When the CP violating interaction ${\cal O}_{PS}$ is switched on}, the DM can annihilate into SM particles without any suppression. Moreover, this interaction is very hard to explore in both DM direct detection and collider experiments.\footnote{The scattering between the DM and a nucleon through ${\cal O}_{PS}$ vanishes in the non-relativistic limit of the DM particle. Moreover, this interaction is also difficult to probe at collider experiments unless it induces an invisible decay of the Higgs boson, namely when $m_{\rm DM} < m_h/2$ ($m_h$ being the Higgs mass).} As a result, when physics of the DM is governed by this CP-violating interaction, only indirect detection experiments allow us to test such a possibility in the near future.

\end{itemize}
DM indirect detection experiments currently do not lead to significant constraints as long as we are considering the singlet-like Majorana fermion DM with the CP-violating interaction absent and assuming no large chirality-flip effect in the dimension-seven operator. We therefore do not include indirect detection constraints in our analysis, though those will be eventually important in the near future.

\subsection{Constraints from collider experiments}

We use simplified models instead of the EFT in order to take into account the present constraints from collider experiments. According to the simplified model Lagrangians discussed in the previous section, we consider the following two models:
\begin{eqnarray}
	&& {\cal L}^{(+)} =
	\sum_f {\cal L}_f^{(+)}[m_{\rm DM}, \Lambda, c_f] +
	{\cal L}_H^{(+)}[m_{\rm DM}, \Lambda, c_H] +
	{\cal L}_S^{(+)}[m_{\rm DM}, \Lambda, c_S],
	\label{eq: UV+}
	\\
	&&	{\cal L}^{(-)} =
	\sum_f {\cal L}_f^{(-)}[m_{\rm DM}, \Lambda, c_f] +
	{\cal L}_H^{(-)}[m_{\rm DM}, \Lambda, c_H] +
	{\cal L}_S^{(-)}[m_{\rm DM}, \Lambda, c_S],
	\label{eq: UV-}
\end{eqnarray}
where the components of the Lagrangians have already been defined in the previous section. Thanks to the mapping between the simplified models and the EFT, both the models are defined using the same parameters as those of the EFT. In the computation of physical quantities associated with collider experiments, we consider these two models separately, namely we define two likelihood functions: one is a function based on the model described by Eq.\,(\ref{eq: UV+}) and the other is described by Eq.(\ref{eq: UV-}).

The collider constraints considered are the invisible decay widths of the Higgs and the $Z$ bosons, the mono-photon and the mono-jet cross-sections. The first and the last ones concern DM searches at the LHC experiment, while the others are from the LEP experiment. The likelihood function $L_{\rm Coll}^{(\pm)}$ is composed of the following four functions:
\begin{eqnarray}
	L_{\rm Coll}^{(\pm)}
	\propto L_{\rm InvH}^{(\pm)}[m_{\rm DM}, \Lambda, \cdots]
	\times L_{\rm InvZ}^{(\pm)}[m_{\rm DM}, \Lambda, \cdots]
	\times L_{\gamma}^{(\pm)}[m_{\rm DM}, \Lambda, \cdots]
	\times L_{\rm Jet}^{(\pm)}[m_{\rm DM}, \Lambda, \cdots],
\end{eqnarray}
where the superscripts $(+)$ and $(-)$ denote likelihood functions constructed based on the models in Eq.\,(\ref{eq: UV+}) and Eq.(\ref{eq: UV-}). We discuss each component in further detail below.

\subsubsection{The likelihood function $L_{\rm InvH}^{(\pm)}$}

When the DM mass is less than a half of the Higgs boson mass, the Higgs boson can decay invisibly into a pair of DM particles via the interactions in ${\cal L}^{(\pm)}_S$. As we have mentioned in section\,\ref{subsec: caveat}, we focus on the case in which the intermediate heavy particles introduced in the simplified models do not lead to any sizeable corrections to usual SM processes. The production cross-section of the Higgs boson as well as its decay into SM particles are thus assumed to be unaltered. Then, a constraint on the invisible decay branching ratio comes from a global fit of the LHC Higgs data under the setup addressed above, which leads to an upper bound on the branching ratio as ${\rm Br}(h \to \chi\chi) \leq 0.24$ at 90\% C.L.\,\cite{Giardino:2013bma}. Since no significant excess from the SM prediction has been observed yet, we adopt the following Gaussian likelihood function for $L_{\rm InvH}^{(\pm)}$ with a mean value of zero:
\begin{eqnarray}
	L_{\rm InvH}^{(\pm)}[m_{\rm DM}, \Lambda, c_S] =
	\exp\left[-\frac{{\rm Br}^{(\pm)}(h \to \chi\chi)^2}
	{2\{\delta{\rm Br}(h \to \chi\chi)\}^2}\right].
\end{eqnarray}
Using the invisible decay width of the Higgs boson, $\Gamma^{(\pm)}(h \to \chi\chi)$, the branching ratio is defined as ${\rm Br}^{(\pm)}(h \to \chi\chi) \equiv \Gamma^{(\pm)}(h \to \chi\chi)/[\Gamma^{\rm (SM)}_h + \Gamma^{(\pm)}(h \to \chi\chi)]$, where the total decay width of the Higgs boson within the SM is denoted by $\Gamma^{\rm (SM)}_h \simeq 4.08$\,MeV\,\cite{Heinemeyer:2013tqa} when $m_h = 125.09$\,GeV\,\cite{Aad:2015zhl}. With $v \simeq 246$\,GeV being the vacuum expectation value of the Higgs field, the simplified models\,(\ref{eq: UV+}) and (\ref{eq: UV-}) predict the widths as
\begin{eqnarray}
	\Gamma^{(+)}(h \to \chi\chi) \simeq
	\Gamma^{(-)}(h \to \chi\chi) \simeq
	\frac{c_S^2 v^2}{16 \pi \Lambda_S^2} m_h
	\left(1 - \frac{4m_{\rm DM}^2}{m_h^2}\right)^{3/2},
	\label{eq: Inv H}
\end{eqnarray}
when $\Lambda_S$ is much larger than the electroweak scale. This result therefore coincides with the one we obtain from EFT, and we shall adopt it in our analysis. The $1\sigma$ uncertainty of the branching fraction is given by $\delta{\rm Br}(h \to \chi\chi) = 0.24/1.64$.

Although the search for the invisible Higgs decay will be improved at on-going and future LHC experiments, thus leading to stronger limits on the coupling $c_S$, SI DM direct detection experiments will provide a more stringent limit on $c_S$ in the near future.

\subsubsection{The likelihood function $L_{\rm InvZ}^{(\pm)}$}

When the DM particle is lighter than a half of the $Z$ boson mass, $Z$ can decay into a pair of DM particles via interactions in ${\cal L}^{(\pm)}_H$. In fact, the decay width of $Z$ has already been measured precisely at the LEP experiment. Apart from the width originating in $Z \to \nu\bar{\nu}$, the upper bound on the invisible decay width of $Z$ is 2\,MeV at 90\% confidence level\,\cite{ALEPH:2005ab}. We therefore consider the following likelihood function for $L_{\rm InvZ}^{(\pm)}$:
\begin{eqnarray}
	L_{\rm InvZ}^{(\pm)}[m_{\rm DM}, \Lambda, c_H] =
	\exp\left[-\frac{\Gamma^{(\pm)}(Z \to \chi\chi)^2}
	{2\{\delta \Gamma(Z \to \chi\chi)\}^2}\right].
\end{eqnarray}
The experimental uncertainty of the invisible decay width $\delta\Gamma(Z \to \chi\chi)$ is 2\,MeV$/1.64$. On the other hand, the simplified models\,(\ref{eq: UV+}) and (\ref{eq: UV-}) predict the width as
\begin{eqnarray}
	\Gamma^{(+)}(Z \to \chi\chi) \simeq \Gamma^{(-)}(Z \to \chi\chi) \simeq
	\frac{c_H^2 v^2 m_Z^3}{32 \pi \Lambda_H^4} \left(1 - \frac{4m_{\rm DM}^2}{m_Z^2}\right)^{3/2},
	\label{eq: InvZ}
\end{eqnarray}
when $\Lambda_H$ is much larger than the electroweak scale. Here $m_Z \simeq 91.2$\,GeV is the $Z$ boson mass. This result again coincides with the one of the EFT and we adopt the invisible decay width\,(\ref{eq: InvZ}) in our analysis.

\subsubsection{The likelihood function $L_{\gamma}^{(\pm)}$}
\label{sec:LEP}

Null results in the search for an excess beyond the SM predictions in single photon events at the LEP2 experiment give an upper limit on the cross-section of the mono-photon process $e^+e^- \to \chi \chi \gamma$. We consider the limit reported by the DELPHI collaboration based on 650\,pb$^{-1}$ data with the centre of mass energy of 180--209\,GeV\,\cite{Abdallah:2003np, Abdallah:2008aa}. The likelihood function $L_{\gamma}^{(\pm)}$ is given by a convolution of Poisson and Gaussian distributions:
\begin{eqnarray}
	L_{\gamma}^{(\pm)}[m_{\rm DM}, \Lambda, c_H, c_L, c_E; B_i] =
	\prod_{i = 1, 2}\,\frac{(S_i^{(\pm)} + B_i')^{N_i} \exp[-(S_i^{(\pm)} + B'_i)]}{N_i!}
	\exp\left[-\frac{(B_i' - B_i)^2}{2\sigma_{B_i}^2}\right],
\end{eqnarray}
where $N_1 = 498$ ($N_2 = 705$) is the number of events observed at the High Density Projection Chamber (the Forward Electro-Magnetic Calorimeter) in the DELPHI detector, which covers the single photon events with the polar angle in the interval $45^\circ \leq \theta \leq 135^\circ$ ($12^\circ \leq \theta \leq 32^\circ$ \& $148^\circ \leq \theta \leq 168^\circ$). On the other hand, $B_1 = 540.6$ ($B_2 = 675.1$) is the expected number of background events with an uncertainty of $\sigma_{B_1} = 4$ ($\sigma_{B_2} = 3$) events. We introduced a nuisance parameter $B_1^\prime$ ($B_2^\prime$) to deal with this uncertainty, which is profiled out by maximizing $L_{\gamma}^{(\pm)}$ in the interval $0 \leq B_1^\prime (B_2^\prime) \leq \infty$.

Using the simplified models\,(\ref{eq: UV+}) and (\ref{eq: UV-}), the expected number of signal events $S_i^{(\pm)}$ are computed using {\tt MadGraph5}\,\cite{Alwall:2011uj} with the model files generated by {\tt FeynRules} and our own detector simulation code modelling the DELPHI detector response.\footnote{The results of the DELPHI collaboration on the dominant SM background process $e^+e^- \to \nu_\ell \bar{\nu}_\ell \gamma$ are reproduced by our simulation code to a good accuracy. Then, the same detector simulation setup is used in our analysis to compute the signal events $S_i^{(\pm)}$ in the likelihood function $L_{\gamma}^{(\pm)}$.} Since the mono-photon process can proceed through the s-channel exchange of the $Z$ boson (from the operator ${\cal L}^{(\pm)}_H$) and through the interactions in ${\cal L}^{(\pm)}_L$ and ${\cal L}^{(\pm)}_E$, the likelihood function $L_{\gamma}^{(\pm)}$ depends on the parameters $c_H$, $c_L$ and $c_E$, in addition to $m_{\rm DM}$ and $\Lambda$.

Our method to match the simplified models onto the effective operators does not fix the total width of the mediator particle, which is therefore an additional free parameter. There is of course a minimum width of the intermediate particles as they necessarily couple to the DM and the SM sectors with $\mathcal{O}(1)$ or smaller coupling factors. In addition, they might also couple to other new states that are not included within the simplified models. Although a larger width typically leads to a reduced signal cross-section\,\cite{Fox:2011pm, Fox:2011fx, Buchmueller:2013dya, Papucci:2014iwa}, having too large a width is not consistent with our basic assumption that physics of the DM is described by a weakly interacting theory. We thus fix the mediator width to be $\Gamma = \Lambda/2$ for our final likelihood scans, which makes the collider constraints conservative, even though for comparison we shall show certain results both for the minimal and maximal width cases.

\subsubsection{The likelihood function $L_{\rm Jet}^{(\pm)}$}

DM particles can be pair-produced at the LHC not only through their dimension-6 couplings with quarks, but also through an intermediate $Z$ or Higgs boson. As is well-known, since the DM itself is invisible, such events are triggered by the presence of at least one hadronic jet, and the events are characterized by the presence of a large missing transverse momentum $\slashed{E}_T$ due to the recoiling DM pair. The construction of the likelihood function $L_{\rm Jet}^{(\pm)}$ is essentially the same as that for the mono-photon process. Both the ATLAS and the CMS collaborations have already searched for such mono-jet events using 7 and 8\,TeV LHC data. In our analysis, we utilize the CMS result based on 19.5\,fb$^{-1}$ data collected during the 8\,TeV run\,\cite{CMS-PAS-EXO-12-048, Khachatryan:2014rra}. Since the data are consistent with the standard model predictions, we adopt the following likelihood function for $L_{\rm Jet}^{(\pm)}$:
\begin{eqnarray}
	L_{\rm Jet}^{(\pm)}[m_{\rm DM}, \Lambda, c_H, c_Q, c_U, c_D; B_3] =
	\frac{(S_3^{(\pm)} + B_3')^{N_3} \exp[-(S_3^{(\pm)} + B'_3)]}{N_3!}
	\exp\left[-\frac{(B_3' - B_3)^2}{2\sigma_{B_3}^2}\right],
\end{eqnarray}
where $N_3 = 3677$ is the number of observed events after employing the kinematical selection criterion used in Ref.\,\cite{CMS-PAS-EXO-12-048}, while the corresponding expected number of background events is $B_3 = 3663.0$ with its uncertainty $\sigma_{B_3} = 196.0$. We have introduced a nuisance parameter $B_3^\prime$ as in the previous case, which is eventually profiled out.

In order to compute the expected number of signal events $S_3^{(\pm)}$ based on the simplified models\,(\ref{eq: UV+}) and (\ref{eq: UV-}), we have performed a Monte Carlo analysis within these frameworks following the {\tt FeynRules}-{\tt MadGraph5}-{\tt Pythia6}\,\cite{Sjostrand:2006za}-{\tt Delphes2}\,\cite{Ovyn:2009tx} chain. We adopt the anti-$k_T$ algorithm\,\cite{Cacciari:2008gp} for jet reconstructions with a cone size of $R = 0.4$ as implemented in the {\tt FastJet2} code\,\cite{Cacciari:2011ma, Cacciari:2005hq}, while {\tt CTEQ6L1}\,\cite{Pumplin:2002vw, Whalley:2005nh} has been used for parton distribution functions with the factorization and the renormalization scales being set at the default dynamical scale choice of the {\tt MadGraph5} code.\footnote{We have reproduced the CMS results (with an agreement to within 5\%) to validate our code by comparing with the expected limit on the DM-quark four-Fermi interactions via an axial vector current.} Since the mono-jet process is from the diagram with the s-channel exchange of the $Z$ boson via interactions in ${\cal L}^{(\pm)}_H$ and diagrams via interactions in ${\cal L}^{(\pm)}_Q$, ${\cal L}^{(\pm)}_U$ and ${\cal L}^{(\pm)}_D$, $L_{\rm Jet}^{(\pm)}$ depends on the parameters $c_H$, $c_Q$, $c_U$ and $c_D$ in addition to $m_{\rm DM}$ and $\Lambda$. The total decay widths of intermediate heavy particles are set to be the same as those for the mono-photon process.

\subsection{Mono-jet bounds recap: effective operators and simplified models}

\begin{figure}[t]
	\includegraphics[width=0.48\textwidth]{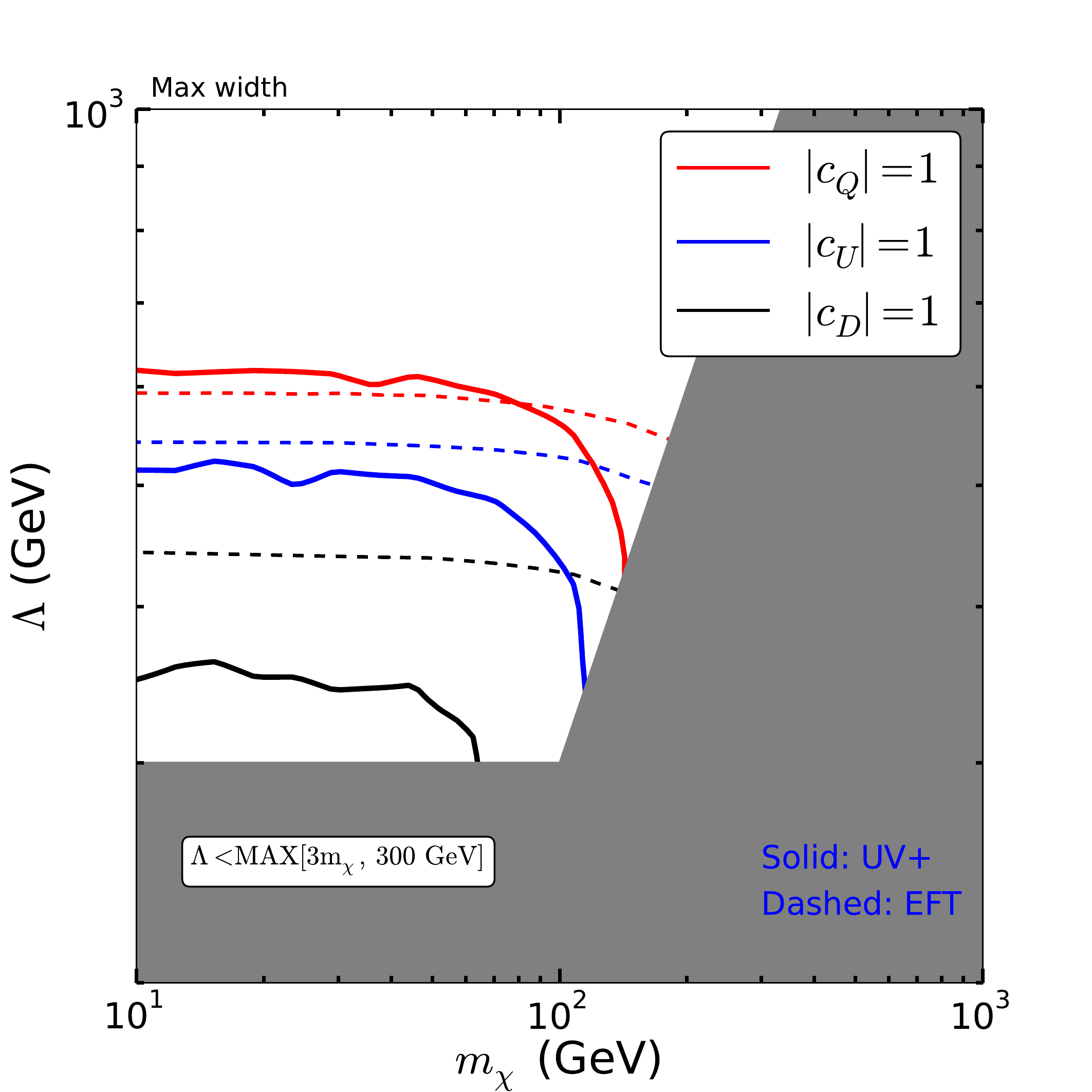}
	\includegraphics[width=0.48\textwidth]{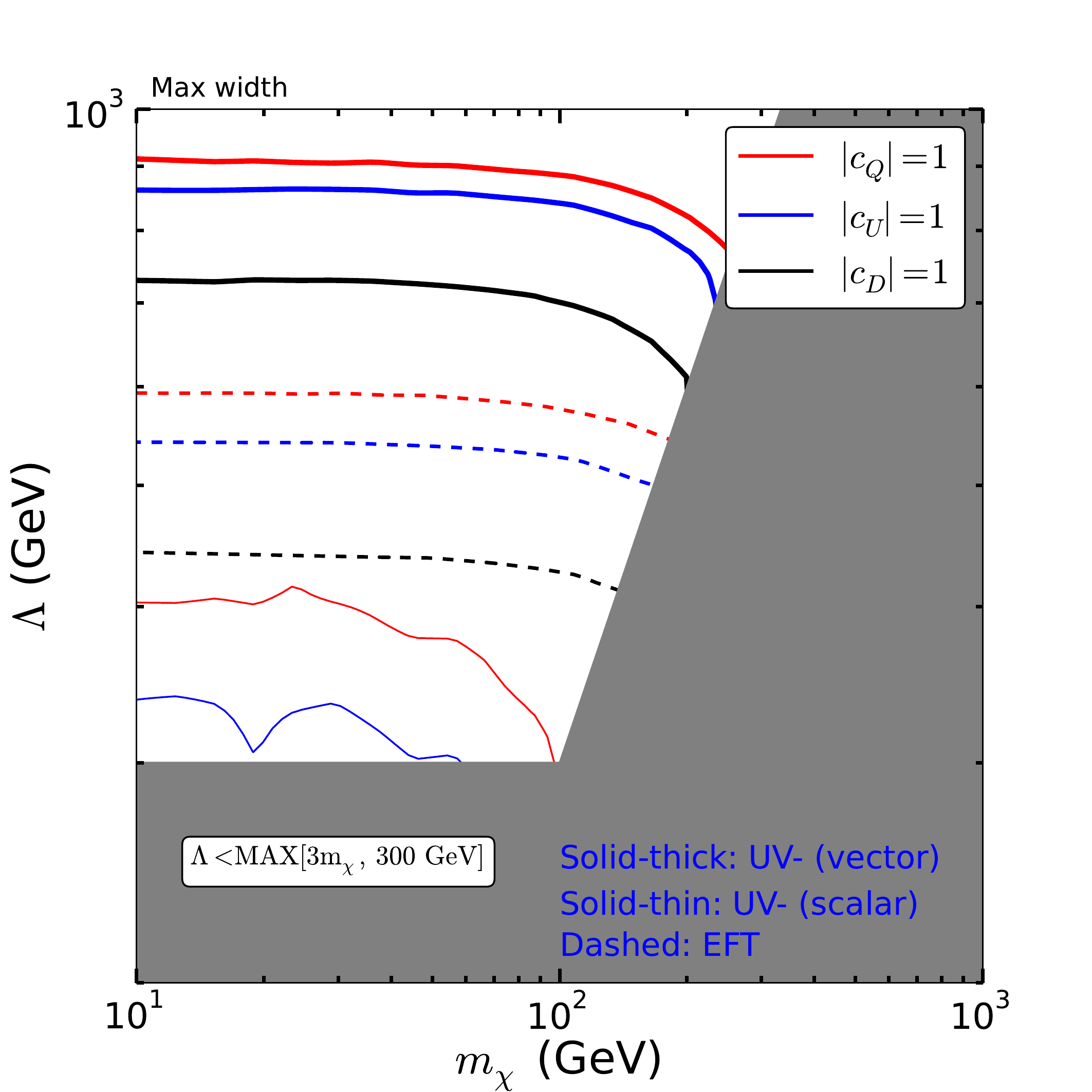}
	\caption{\sl \small Comparison between the $95\%$ confidence level exclusion contours obtained using the EFT and different simplified models, based on the mono-jet and missing energy searches at the 8 TeV LHC, with $Z_2$-even mediator models (left panel) and scalar or vector $Z_2$-odd mediator models (right panel). The width of the mediators have been fixed to be half of the mediator mass, which is the maximum value considered in our analysis.}
	\label{fig: EFT_SM}
\end{figure}

The comparison of the 8 TeV LHC bounds described in the previous subsection on the effective scale $\Lambda$ (or equivalently mediator masses) as a function of the DM mass, obtained using effective operators or simplified models, has been discussed extensively in the literature, and we refer the reader to Refs.\cite{Buchmueller:2013dya,Busoni:2013lha,Busoni:2014sya,Busoni:2014haa} for details. In this subsection, we just want to note two points important for our study, as illustrated in Fig.\,\ref{fig: EFT_SM}:
\begin{itemize}
	\setlength{\itemsep}{0mm}

	\item Depending upon the UV completion and the particular operator under study, the EFT bounds can be similar, below or above the simplified model bounds across the whole DM mass range (in the region where $\Lambda> 3m_\chi$ with the mediator width fixed to half its mass). Thus, the EFT results cannot be conservative lower estimates for all possible models even for the restricted range of $\Lambda$ considered, and we must resort to simplified models for consistent prediction of collider bounds.

	\item The bounds obtained using the vector mediator in the simplified model\,(\ref{eq: UV+}), the scalar or the vector mediators in the simplified model\,(\ref{eq: UV-}) are different. For each operator case ${\cal O}_Q$, ${\cal O}_U$ and ${\cal O}_D$, we analyze the likelihoods and present the results in Fig.\,\ref{fig: EFT_SM} using these three mediators separately. In the simplified model\,(\ref{eq: UV-}), the vector mediators in general lead to larger DM pair production cross-sections compared to the scalar mediators (for the same mediator and dark matter masses). The bounds are very different between the simplified models\,(\ref{eq: UV+}) and (\ref{eq: UV-}), and this is the reason we present our results in later sections using the two models separately.

\end{itemize}

\subsection{Direct production of the mediators at the LHC}

Compared to the mono-jet and missing energy signal from DM pair production along with a jet, single or pair production of the mediators could be easier to search for at the LHC, especially if the mediators can be produced on-shell. However, there is a degree of model dependence in comparing the monojet signal with the direct production processes. We discuss the relative importance of each in the following.

\subsubsection{Direct production of $Z_2$-even mediator}

In the s-channel UV completion with a $Z^\prime$, the $Z^\prime$ could be looked for as a di-lepton or di-jet resonance, the former case having a significantly lower SM background. As argued in Sec.\,\ref{subsec: caveat}, there could be a consistent UV completion with a $Z^\prime$ that couples only to quarks, in which case the di-jet resonance search will be the only option. Such a search depends strongly on the $Z^\prime$ width $\Gamma_{Z^\prime}$, as illustrated by the CMS search for wide resonances\,\cite{Khachatryan:2015sja}. With $\Lambda$ being the mass of the $Z^\prime$, we gradually lose sensitivity as $\Gamma_{Z^\prime}/\Lambda$ approaches values as high as $0.5$, since an excess over a very broad invariant mass range can be hard to disentangle from QCD backgrounds. Since a large number of degrees of freedom can couple to the $Z^\prime$ (including all the SM quarks and the DM pair), its width can easily be high enough making the resonance searches harder.

\begin{figure}[t]
	\includegraphics[width=0.48\textwidth]{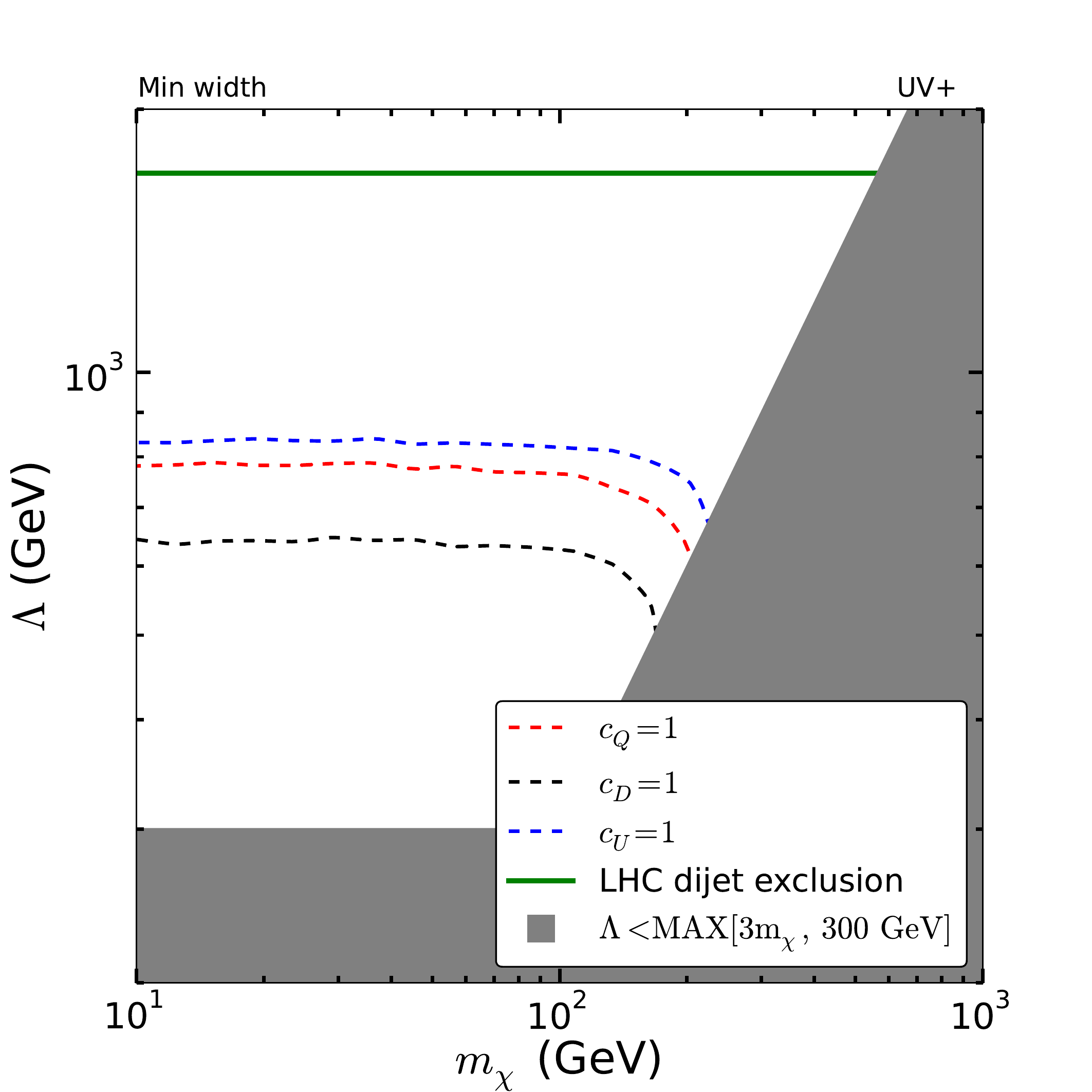}
	\includegraphics[width=0.48\textwidth]{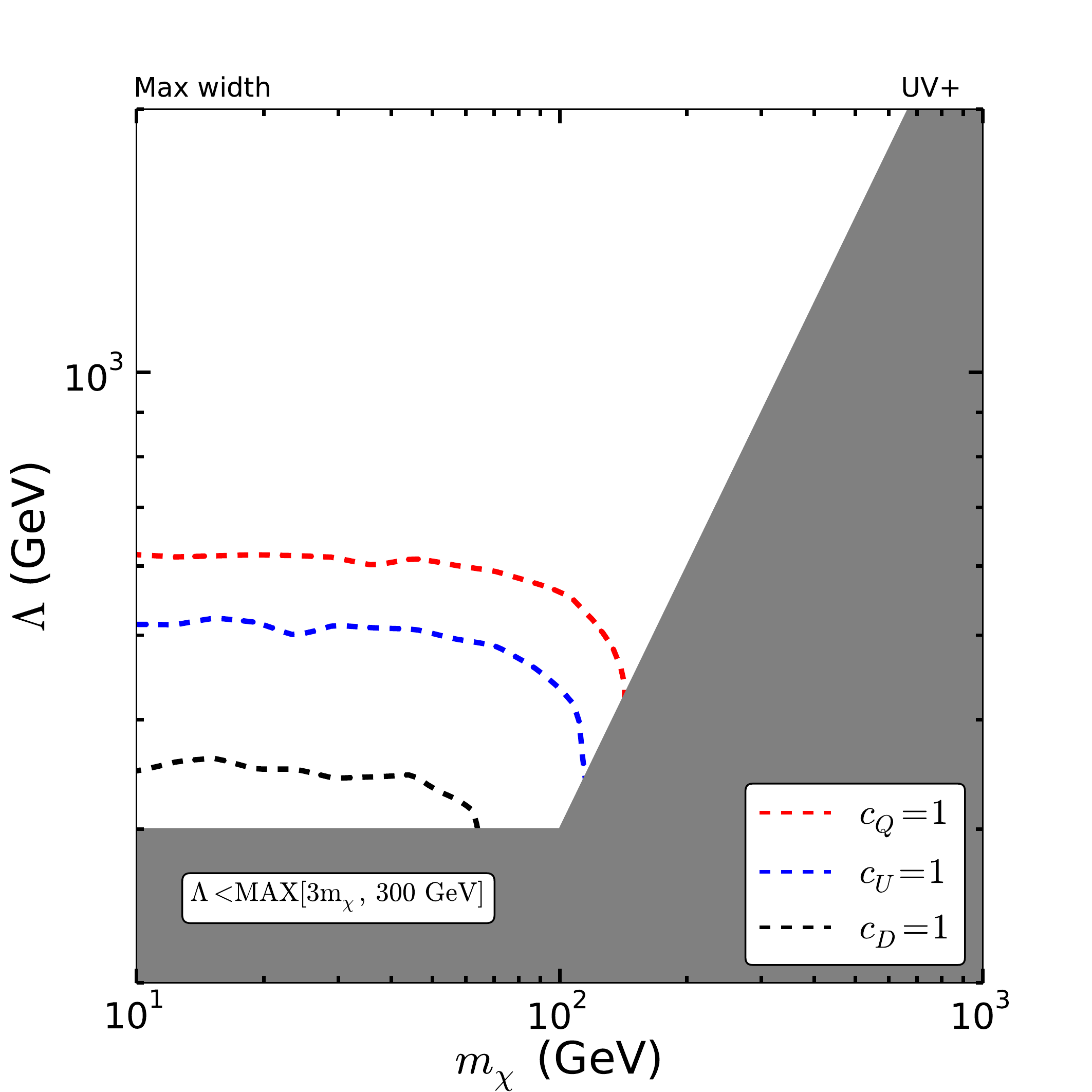}
	\caption{\sl \small Comparison of LHC8 mono-jet (dashed lines) and di-jet (solid line) constraints in the simplified model with a $Z^\prime$ mediator, in the DM mass and mediator mass plane. The left panel shows the constraints for a minimal width of the $Z^\prime$, while the right one for $\Gamma_{Z^\prime} = \Lambda/2$, see text for details. In the latter case the di-jet bounds become much weaker, and are not visible in this panel.}
	\label{fig: uvp_dijet}
\end{figure}

In Fig.\,\ref{fig: uvp_dijet} we compare the LHC8 di-jet (green solid line) vs monojet constraints (dashed blue, red and black lines for different couplings) in the dark matter mass and mediator mass plane by fixing the couplings to one. The left panel is with the minimal width of the $Z^\prime$, i.e., the sum of the partial widths to SM quarks and the DM pair with all couplings set to one. On the right panel we show results with $\Gamma_{Z^\prime} = \Lambda/2$, which is the largest width considered in this study. Unfortunately, for such a large width the di-jet constraint (extrapolated from CMS results in Ref\,\cite{Khachatryan:2015sja}) disappear from the panel, and only the monojet exclusions remain, even though they also become considerably weaker compared to the minimal width scenario. Because of this strong model dependence of the di-jet bounds, it would be a conservative choice not to include them in our general analysis, with the clarification that they do imply strong constraints for specific models. To put it another way, to find viable parameter regions, we need to restrict ourselves to regions where the di-jet constraints are weak, and the large $Z^\prime$ width case is precisely that. 

\subsubsection{Direct production of coloured $Z_2$-odd mediator}

Pair production of $Z_2$ odd coloured mediators (vector or scalar) can lead to multi-jets and missing momentum signals with large QCD cross-sections. The LHC search results for first two generation squarks in the MSSM~\cite{Aad:2014wea} can then be recast to our scenario to obtain the relevant bounds. Once again, the matching adopted by us does not fix the width of the mediator, which introduces an additional model dependence here as well. Let us recall that the vector mediator production leads to a larger cross-section compared to the scalar case, and thus it is sufficient to compare the mono-jet and di-jet bounds for the vector mediators.

\begin{figure}[t]
	\includegraphics[width=0.48\textwidth]{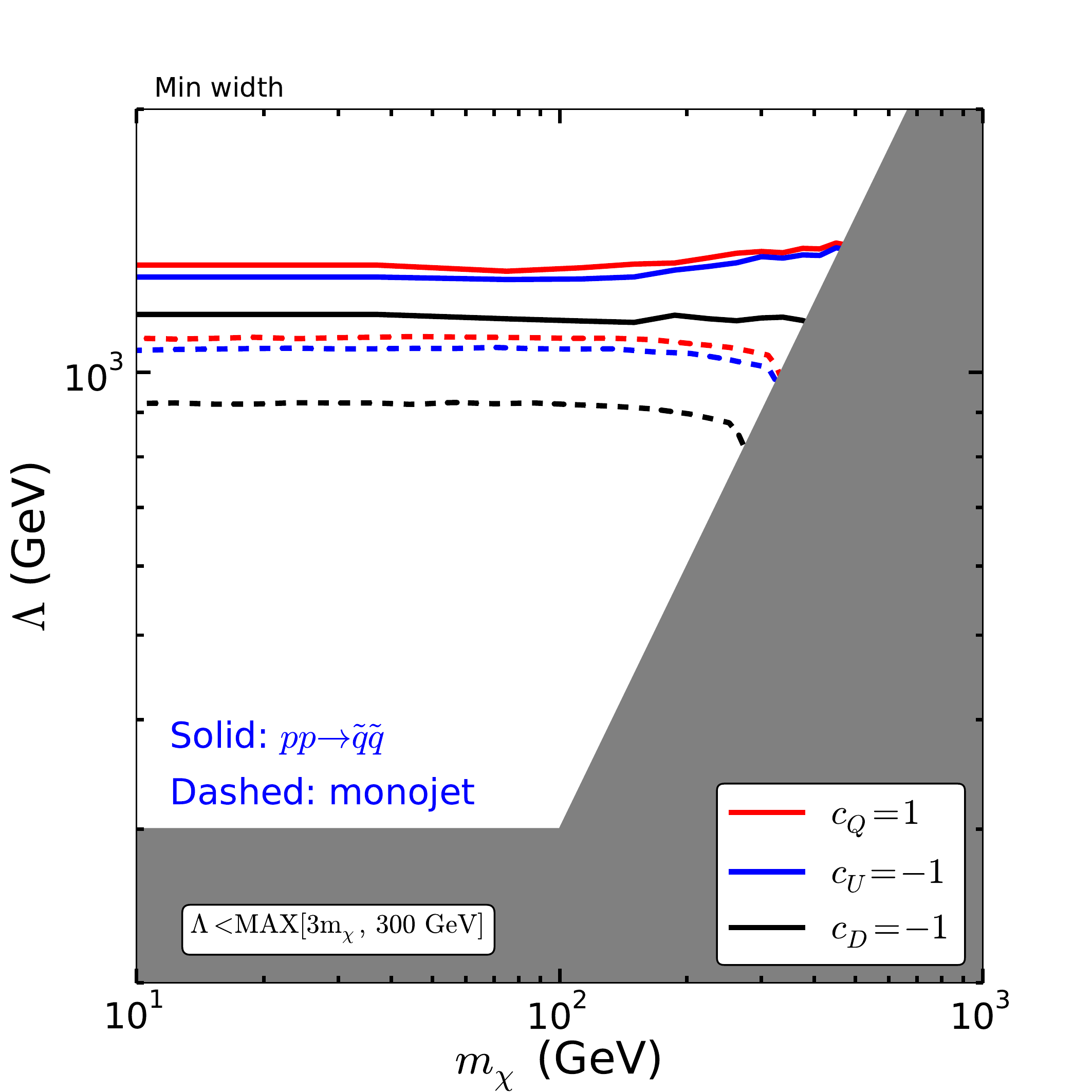}
	\includegraphics[width=0.48\textwidth]{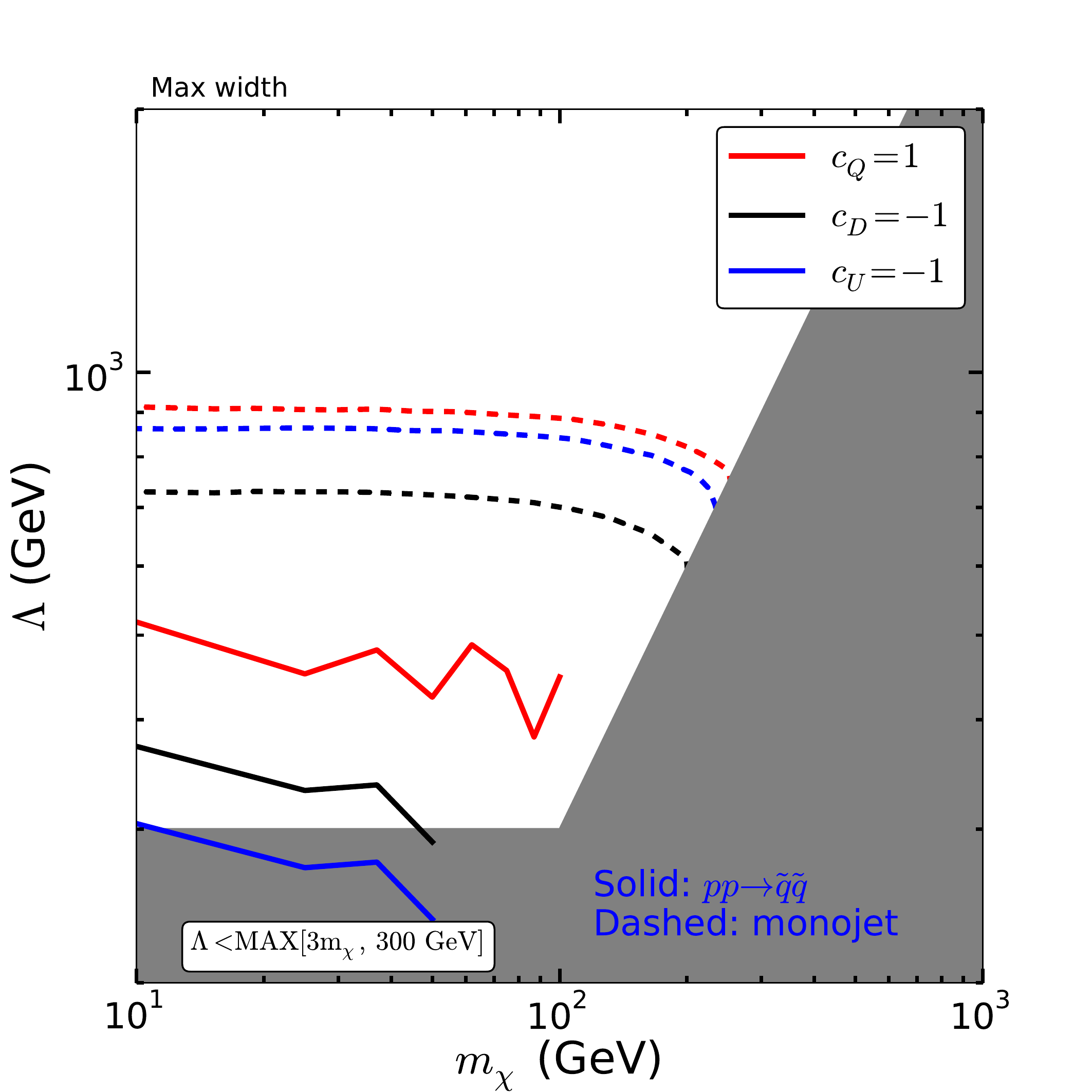}
	\caption{\sl \small Comparison of LHC8 mono-jet and di-jet plus missing momentum constraints in the simplified model with a coloured vector mediator, in the DM mass and mediator mass plane. The left panel shows the constraints for a minimal width of the vector mediator, while the right one for $\Gamma_{\tilde{q}} = \Lambda/2$. See text for details.}
\label{fig: uvm_dijet}
\end{figure}

We show the results in Fig.\,\ref{fig: uvm_dijet}, with the left panel for the minimal width and the right one for $\Gamma_{\tilde{q}} = \Lambda/2$, as before, with $\Gamma_{\tilde{q}}$ and $\Lambda$ being the width and the mass of the $Z_2$-odd mediator. Once again, for the minimal width case, the bounds from direct production of mediators followed by their decay to a DM and a quark, are stronger. However, the mono-jet + missing energy bounds are more stringent if the width of the mediator is very large due to unknown decay modes, simply because they do not decay often enough to final states with high transverse momentum quarks. Therefore, in order to set the most conservative bounds, we focus on scenarios with large mediator widths, and include only the monojet constraints in our final results.

\subsection{Likelihood analysis results: allowed region in the $(m_\chi, \Lambda)$-plane}

We are now in a position to discuss our results of the allowed parameter regions for the singlet-like Majorana fermion DM, after performing the profile-likelihood analysis with all available constraints. As discussed before, since the LHC and LEP bounds differ for the s-channel and t-channel mediator models, we performed separate likelihood analyses in these two cases. The most relevant projection for the $68\%$ and $95\%$ C.L. allowed regions would be in the $(m_\chi, \Lambda)$-plane, which we show in Fig.\,\ref{fig: mx_lam}, with the left panel for the simplified model\,(\ref{eq: UV+}) (called the ${\rm UV}_+$ case hereafter), and the right panel for the simplified model\,(\ref{eq: UV-}) (called the ${\rm UV}_-$ case). The difference between the two models is barely observable in this two dimensional parameter space (although small differences exist in these plots for $m_\chi>100$ GeV). A few points are worth noting:

\begin{figure}[t]
	\includegraphics[width=0.48\textwidth]{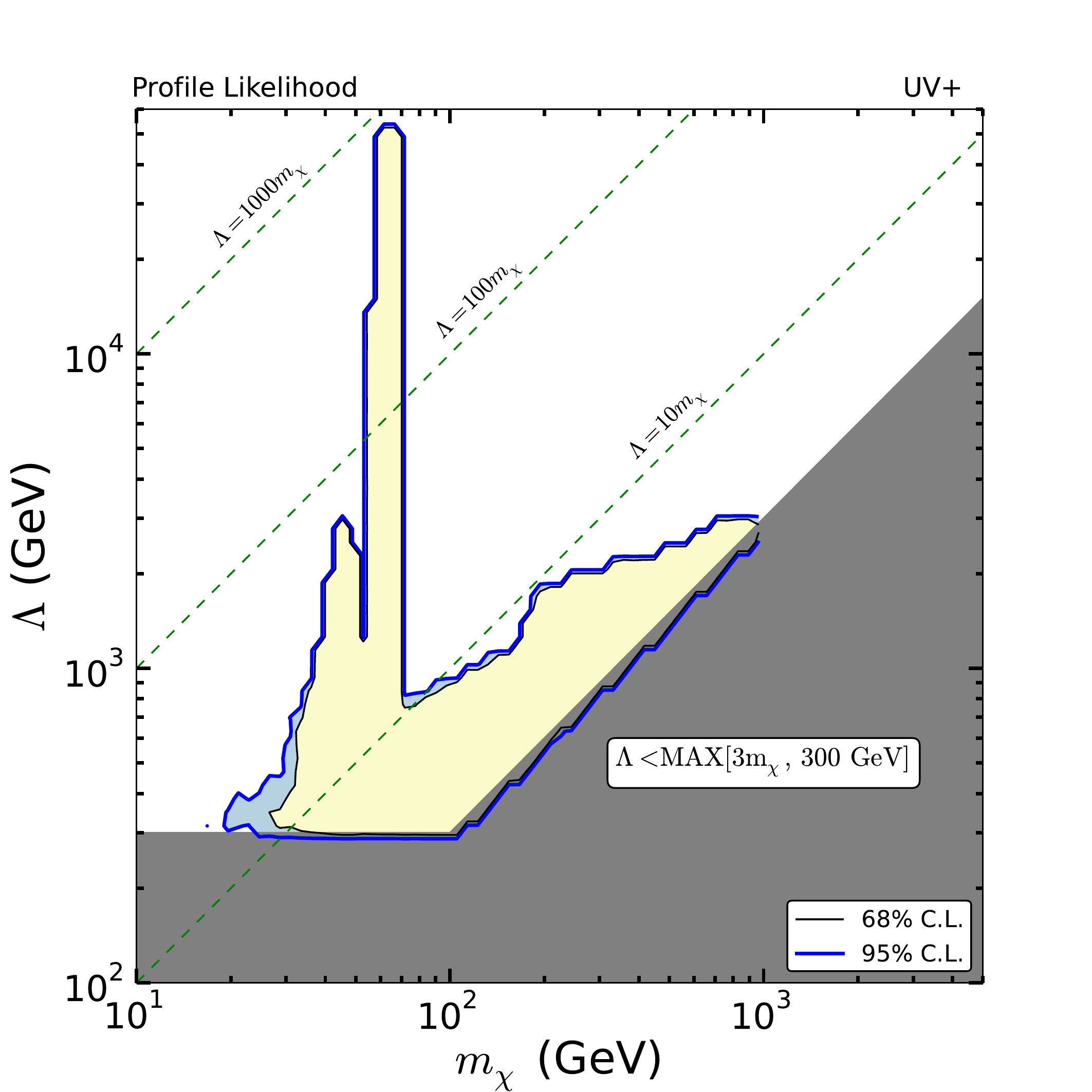}
	\includegraphics[width=0.48\textwidth]{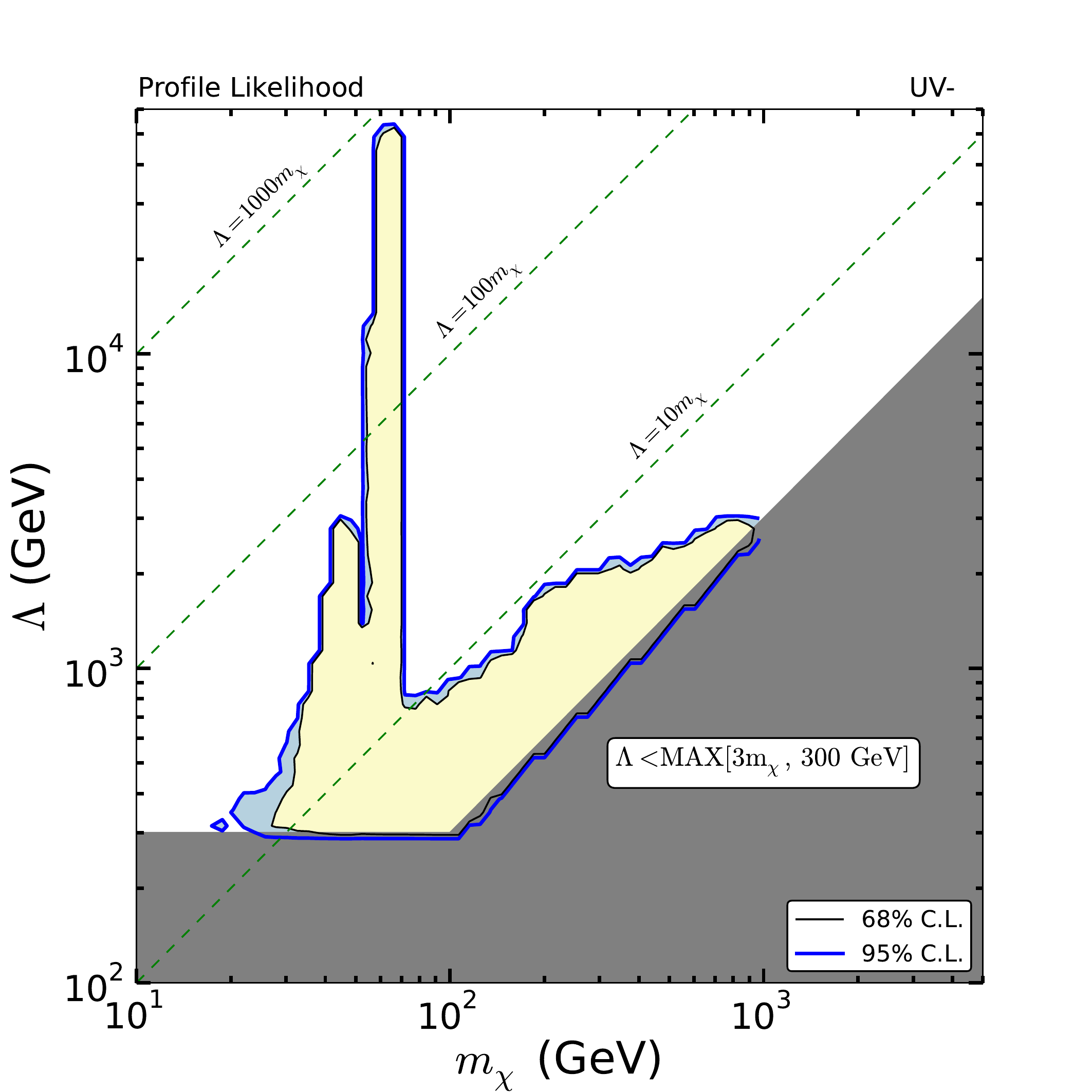}
	\caption{\sl \small Allowed regions at $68\%$ (yellow) and $95\%$ C.L. (blue) in the DM mass ($m_\chi$) and EFT cut-off ($\Lambda$) plane, for the ${\rm UV}_+$ model (left panel) and ${\rm UV}_-$ model (right panel). In the grey shaded region an EFT framework is not suitable for the analysis.}
	\label{fig: mx_lam}
\end{figure}

\begin{enumerate}
\setlength{\itemsep}{0mm}

	\item As discussed at length in our previous study\,\cite{Matsumoto:2014rxa}, the shape of the allowed region is essentially determined by the relic abundance requirement, especially the rather sharply defined upper contour determining the maximum allowed values of $\Lambda$. It is also interesting to note that in the log scale plot, the difference between the $68\%$ and $95\%$ C.L. allowed regions is rather small, since the likelihood is very sensitive to DM annihilation cross-sections, which in turn changes very rapidly with $\Lambda$ (and also with $m_\chi$ in the resonance regions).

	\item The low DM mass region below $20$ GeV is disfavoured, as long as $\Lambda>300$ GeV. It should be noted that in special cases mediators lighter than $300$ GeV might evade all current bounds, thereby opening up this region. However, to properly analyze such cases, we would require a specific model framework.

	\item Unless we are near a resonance for DM pair annihilation, generically the maximum allowed value of $\Lambda$ is around $10\,m_\chi$. On the other hand, near the Z-pole, it extends up to almost $100\,m_\chi$. Due to the very narrow width of the Higgs boson, near the Higgs pole, even $\Lambda=1000\,m_\chi$ is allowed.

	\item The primary reason for the similarity between the ${\rm UV}_+$ and ${\rm UV}_-$ cases is as follows: even though the LHC8 constraints rule out a considerably larger range of values of the quark couplings for a given $\Lambda$ in the ${\rm UV}_-$ case compared to ${\rm UV}_+$, the lepton couplings in the larger DM mass range remain unconstrained. It is the lepton couplings which can compensate for the quark couplings to give the required relic abundance, thereby leading to a very similar allowed $(m_\chi, \Lambda)$-plane for the two cases. 

	\item Beyond the DM mass of 1\,TeV, we clearly do not have any allowed region as long as $\Lambda > 3\,m_\chi$, essentially because the most important annihilation mode in this region to top quark pairs has an s-wave cross-section which is independent of the DM mass and goes as $m_t^2/\Lambda^4$, thereby reducing very rapidly the annihilation cross-section at a rate approximately larger than $m_t^2/(81 m_\chi^4)$.

\end{enumerate}

\begin{figure}[p!]
	\begin{center}
	\includegraphics[width=0.45\textwidth]{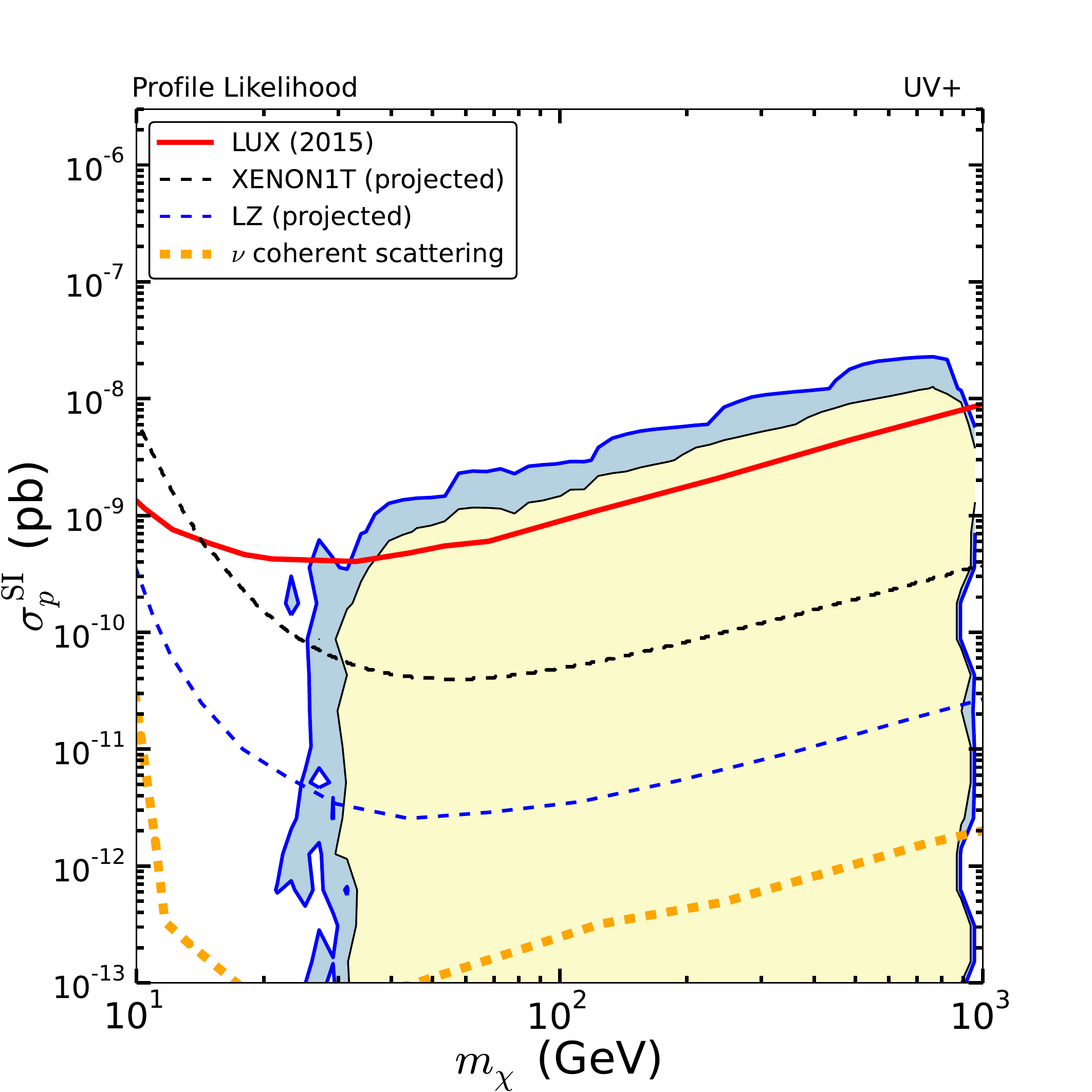}
	\includegraphics[width=0.45\textwidth]{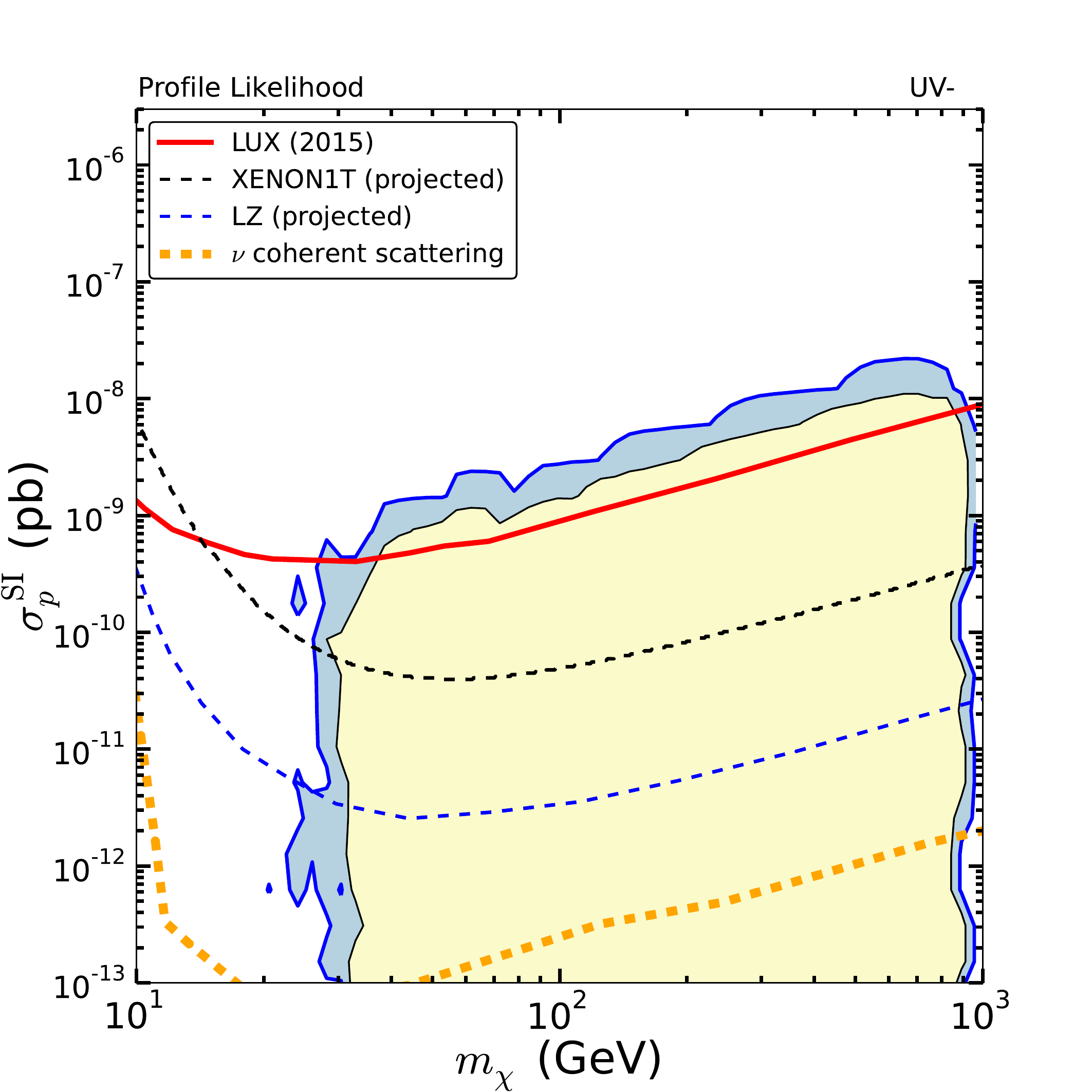} \\
	\includegraphics[width=0.45\textwidth]{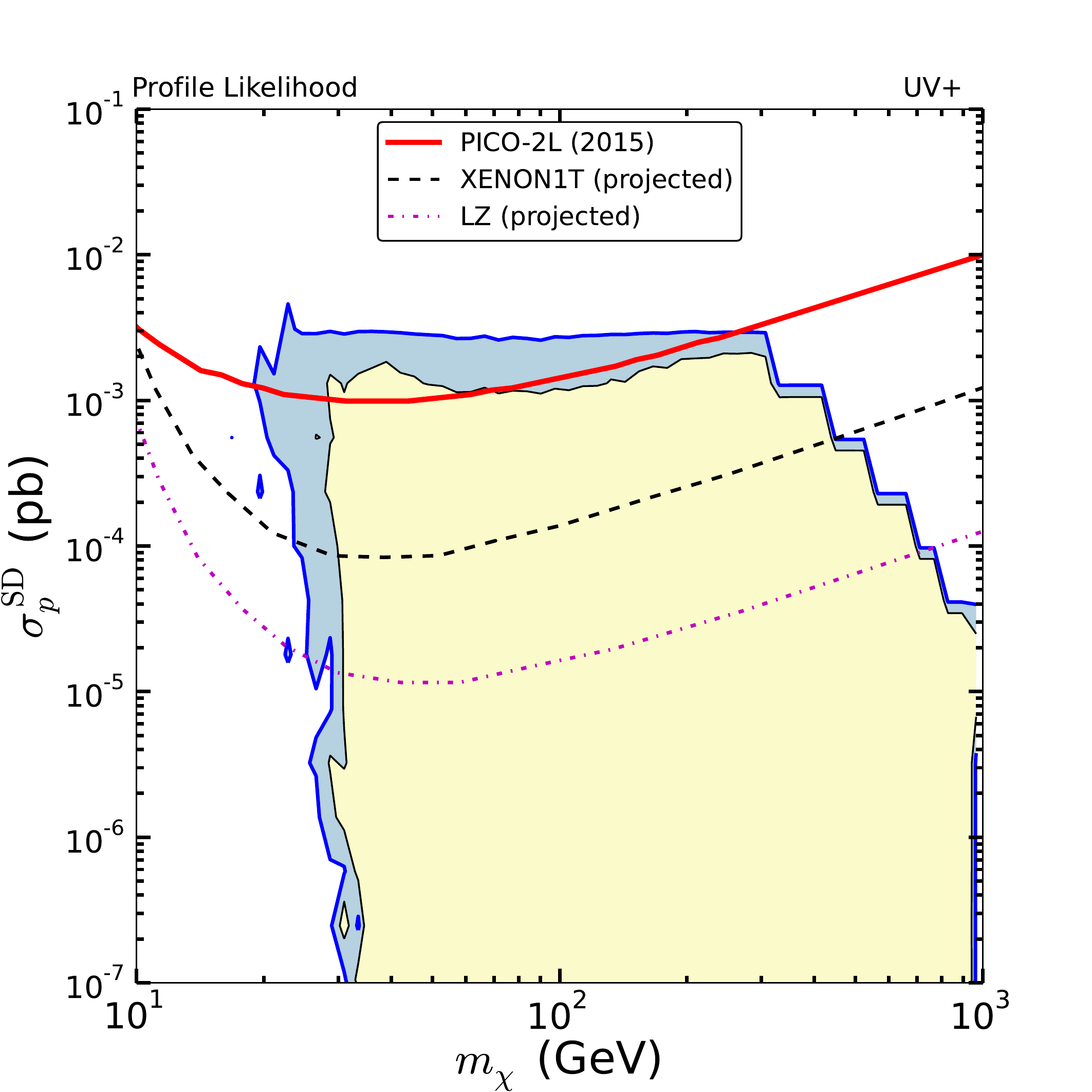}
	\includegraphics[width=0.45\textwidth]{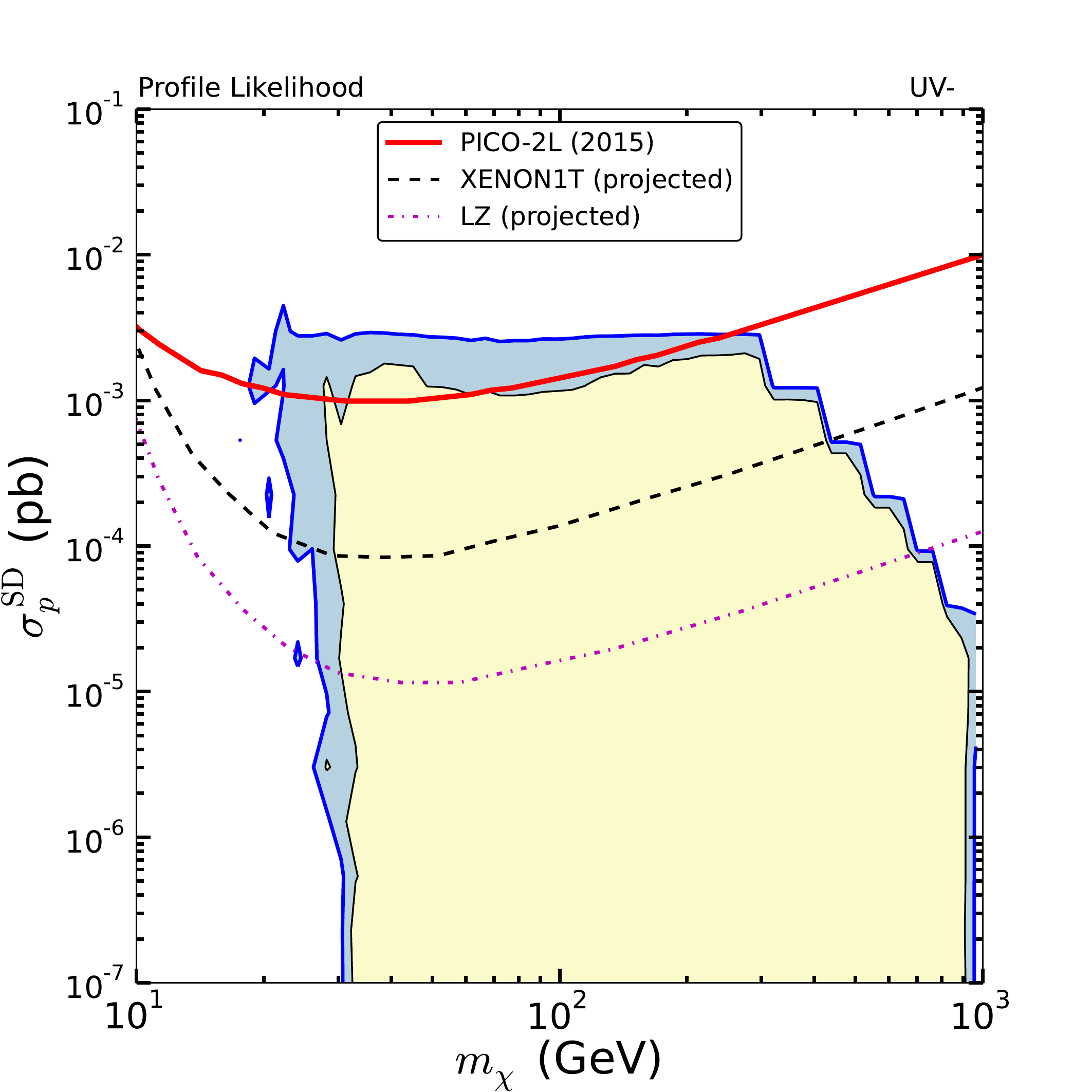} \\
	\includegraphics[width=0.45\textwidth]{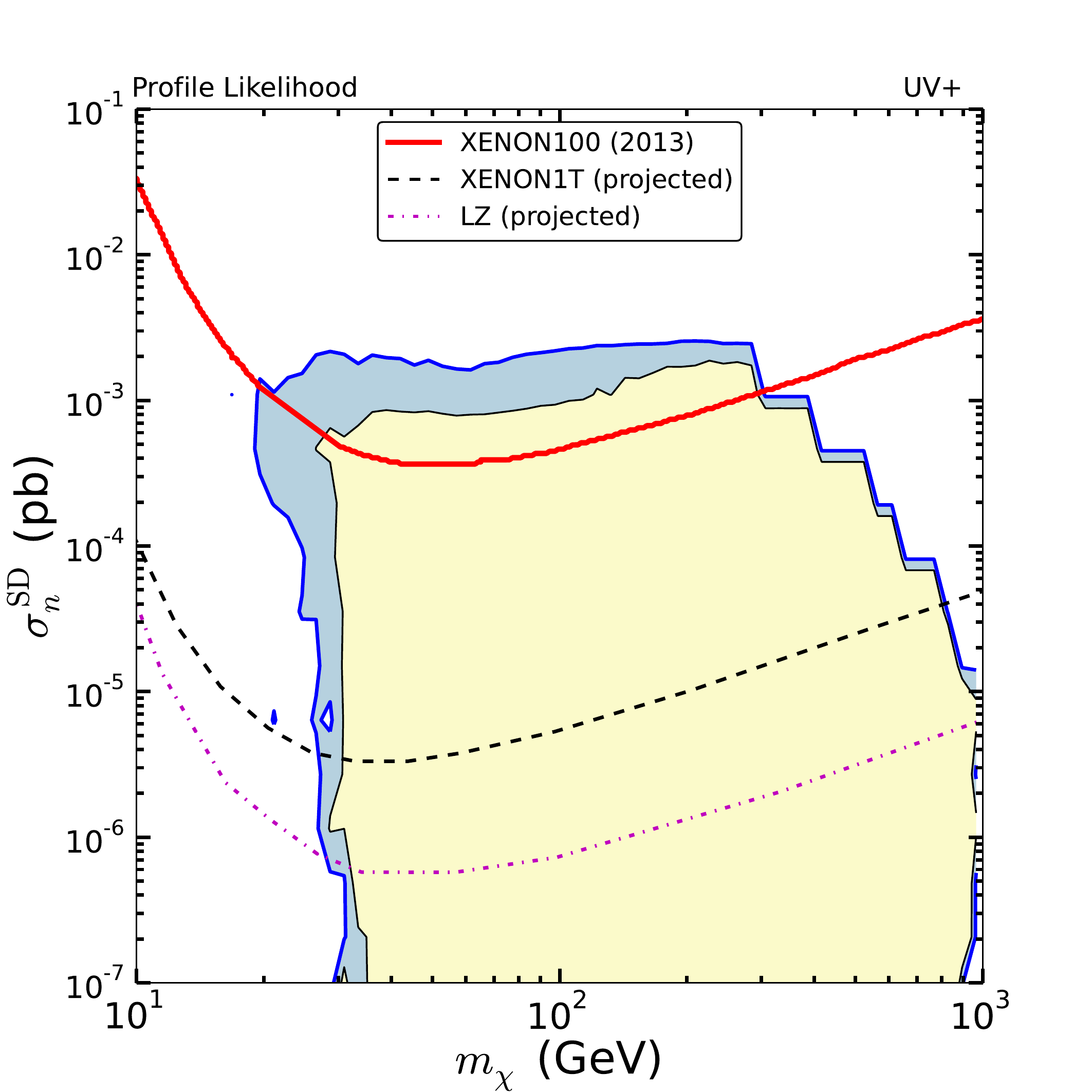}
	\includegraphics[width=0.45\textwidth]{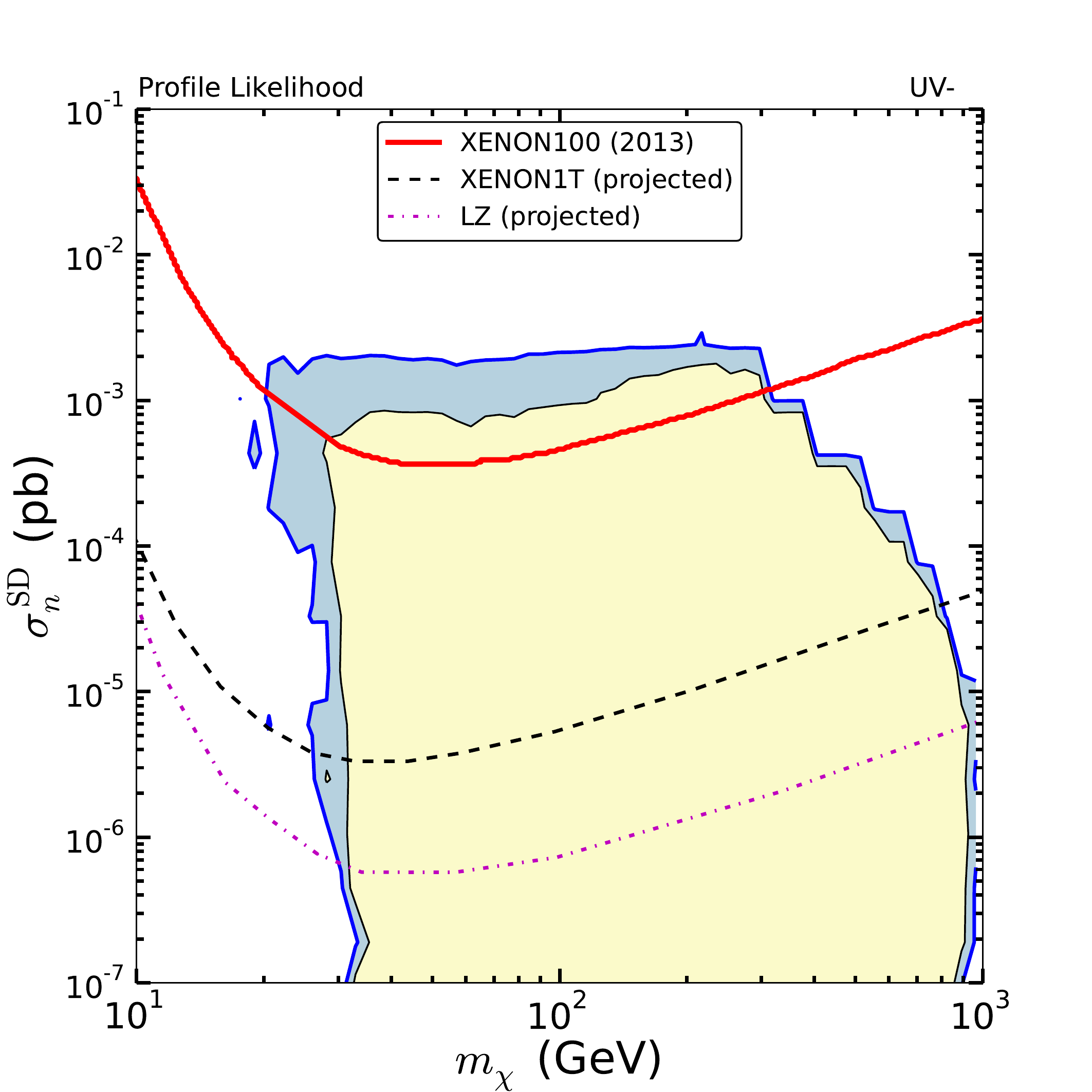}
	\caption{\sl \small Predictions for spin-independent direct detection cross-sections ($\sigma_p^{\rm SI}$) and spin-dependent direct detection cross-sections with proton ($\sigma_p^{\rm SD}$) and neutron ($\sigma_n^{\rm SI}$) in the $68\%$ (yellow) and $95\%$ (blue) allowed region of parameter space for the singlet-like Majorana fermion DM. Comparison with future projections in the XENON1T and LZ experiments are also shown. Predictions in the UV$_+$ case are shown in the three left panels, while the three right panels are for the UV$_-$ case.}
	\label{fig: direct}
	\end{center}
\end{figure}

For a better understanding about the range of scattering cross-sections that need to be probed for this scenario, we show in Fig.\,\ref{fig: direct} the range of SI and SD scattering rates predicted in the currently allowed parameter space. The predictions in the UV$_+$ case are shown in the three left panels, while the three right panels are for the UV$_-$ case. We also display in the same panels the projected reach of future ton-scale direct detection experiments, namely the XENON1T\,\cite{Aprile:2012zx} and LZ\,\cite{Malling:2011va} experiments. The colour coding is the same as before, with yellow and blue corresponding to the $68\%$ and $95\%$ C.L. allowed regions. There is a small mismatch between our allowed regions and the current LUX, XENON100 or PICO-2L exclusions for SI (proton), SD (neutron) and SD (proton) cross-sections respectively. This is primarily because of systematic errors associated with the astrophysical and nuclear physics uncertainties included in our likelihood. It should also be noted that as far as the SI and SD rates are concerned, the UV$_+$ and UV$_-$ models predict the same rates, since they both reduce to the same effective operators for low momentum transfers, and the allowed regions in the $(m_\chi, \Lambda)$-plane, as seen before, are almost identical.

\section{Summary and future prospects}
\label{sec: summary}

To summarize, our goal in this study has been to estimate the mass-scale of the mediator particles allowed by relic density requirements and current constraints on a Majorana fermion WIMP candidate, focusing on weakly coupled models. In order to compute all relevant DM related observables, using a complete set of gauge-invariant effective operators is the most general model independent approach, and as in our previous work\,\cite{Matsumoto:2014rxa} we have adopted the same here. However, such an approach is of limited validity at high-energy colliders, and hence using simplified models to interpret the collider bounds allows us to survey a larger range of the mediator masses. To this end, we first write down all possible simplified models, which lead to each effective operator, when the heavy fields compared to the DM mass and the electroweak scale are integrated out. Matching the simplified model to the EFT above the electroweak scale then gives us a one-to-one map between the EFT parameters and the simplified model ones. We utilize this map to compute the collider observables within the context of the simplified models, and then combine them with the other constraints to obtain the complete likelihood function. Using a profile-likelihood approach we are then able to determine the allowed ranges for the dark matter mass and the EFT cut-off scale, while maximizing the likelihood over rest of the parameters. 

The main conclusions that can be drawn from this analysis are that for $\Lambda > 300$\,GeV, DM masses below $20$\,GeV are disfavoured. The Z- and Higgs-pole regions survive all current constraints up to $\Lambda$ as high as $100\,m_\chi$ and $1000\,m_\chi$, respectively. Beyond the Higgs-pole region for DM heavier than about 70\,GeV and lighter than 1\,TeV, values of the suppression scale up to $\Lambda = 10\,m_\chi$ are viable. The UV$_+$ and UV$_-$ models lead to very similar results, as even though the LHC constraints act differently on the quark couplings' space of the models, the yet unconstrained leptonic couplings at higher DM masses are sufficient to furnish the correct relic abundance.

The next question to ask then would be how much of the remaining space of parameters can the next generation direct detection and planned collider experiments probe? Given the remaining regions explained above, the most important future searches for the singlet-like Majorana fermion DM will be multi-ton scale direct detection experiments (for example, the LZ experiment\,\cite{Malling:2011va}, which is an upgrade of LUX) and proposed future lepton colliders. For illustration, we perform order of magnitude estimates for the reach of $e^+ e^-$ colliders with centre of mass energies of 250\,GeV (which can be reached by both the CEPC\,\cite{CEPC} and the ILC\,\cite{Baer:2013cma}) and an energy of 500\,GeV (which can be achieved by the ILC). A further Giga-$Z$ option of having collisions at an energy equal to $m_Z$ is also under study, and can help in exploring the Z-pole region of DM annihilation further, but we do not include this option in our estimates. In that sense, the major role of the lepton colliders will be to examine the DM coupling to the Z-boson and also the four-fermi interaction with leptons using the mono-photon process. Such processes can of course be relevant only up to the kinematic threshold of DM pair production, which will be 250\,GeV for the 500\,GeV ILC. For our simple estimate we compute the mono-photon cross-sections with basic selection cuts as described before for LEP\,\cite{Abdallah:2003np,Abdallah:2008aa} in Sec.~\ref{sec:LEP}, and compare them to the baseline value of 1\,fb for the $500$ GeV ILC. As has been shown in Ref.\,\cite{Bartels:2012ex}, a signal cross-section of order 1\,fb can be excluded at $95\%$ C.L. for the nominal luminosities and beam polarizations planned for this experiment. 

\begin{figure}[t]
	\includegraphics[width=0.48\textwidth]{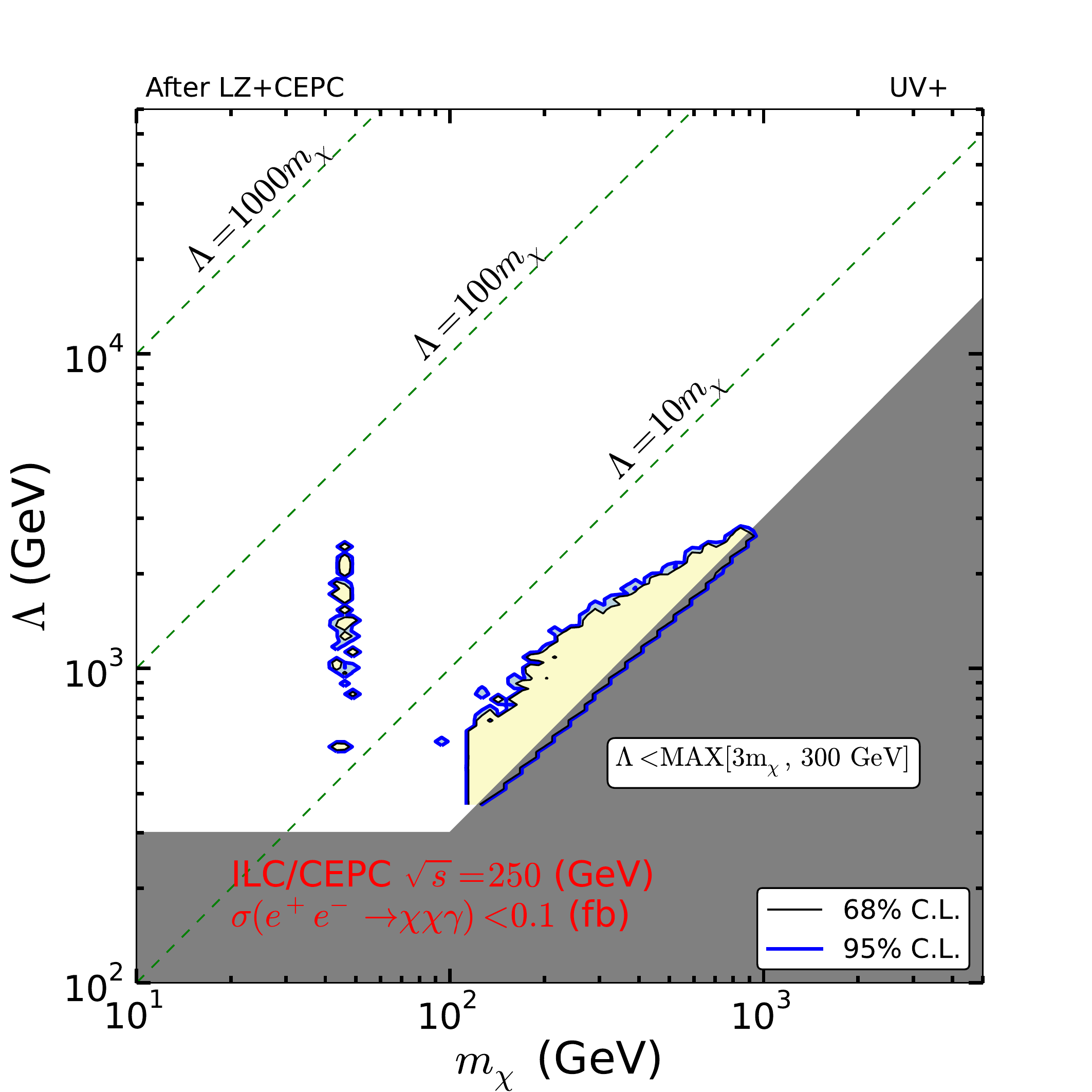}
	\includegraphics[width=0.48\textwidth]{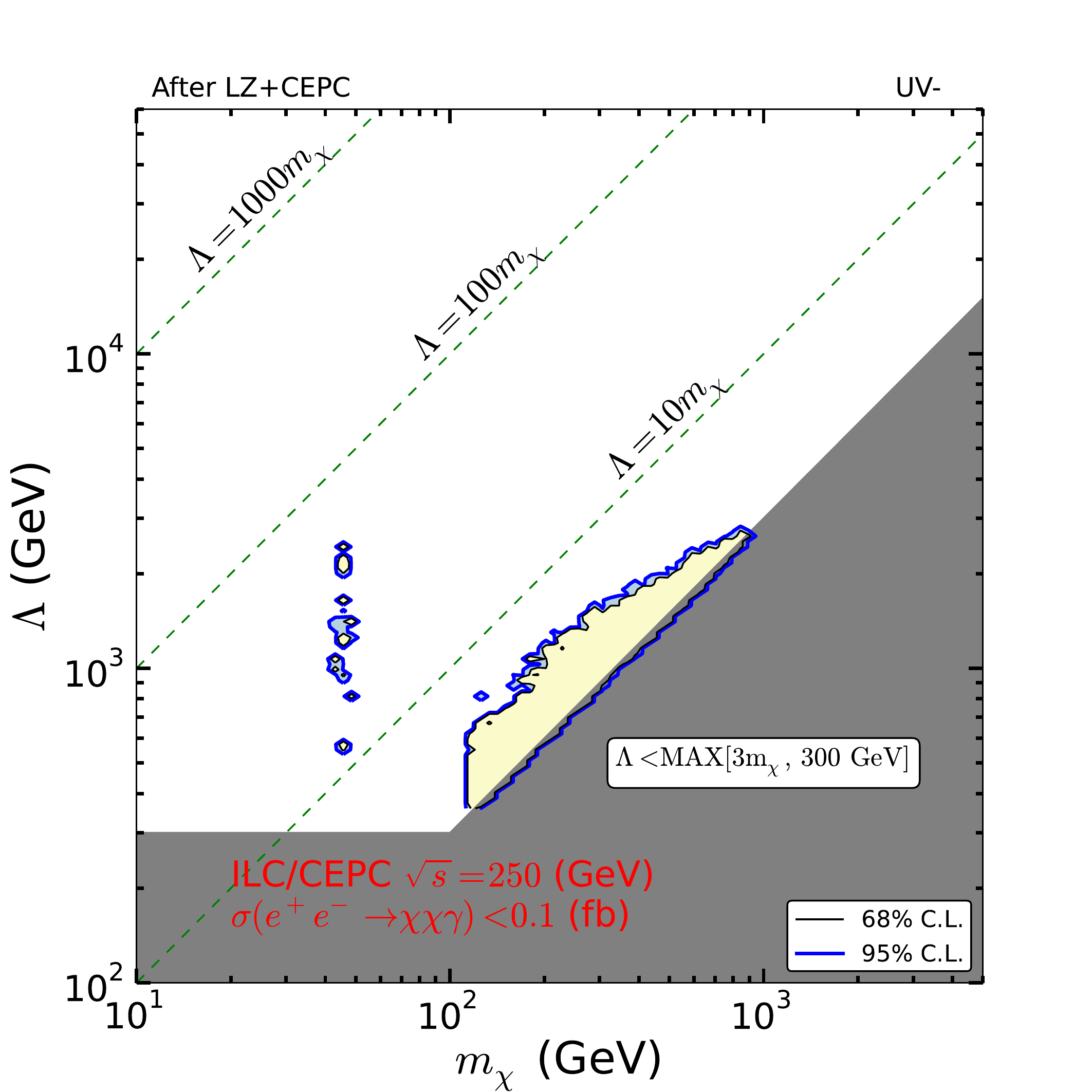} \\
	\includegraphics[width=0.48\textwidth]{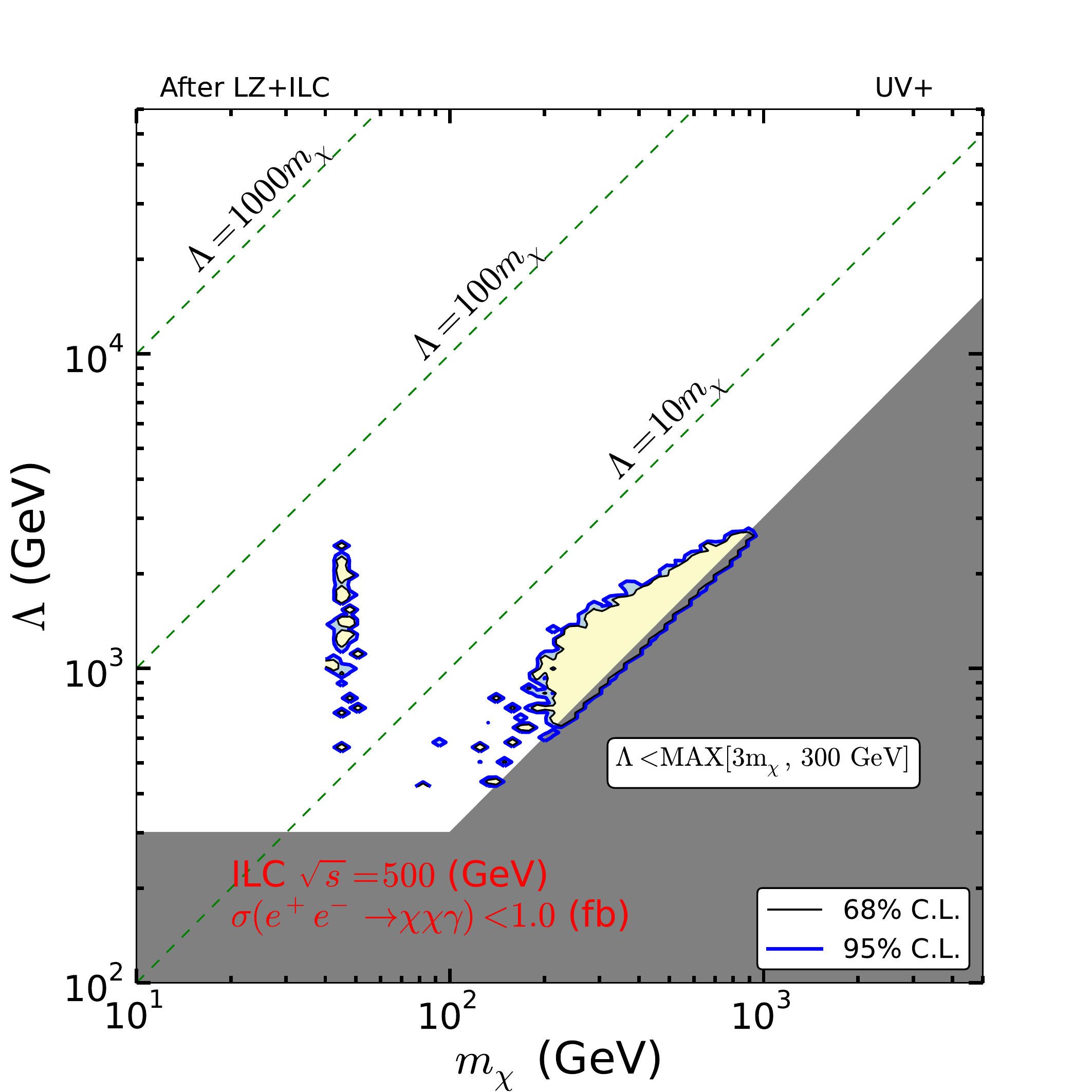}
	\includegraphics[width=0.48\textwidth]{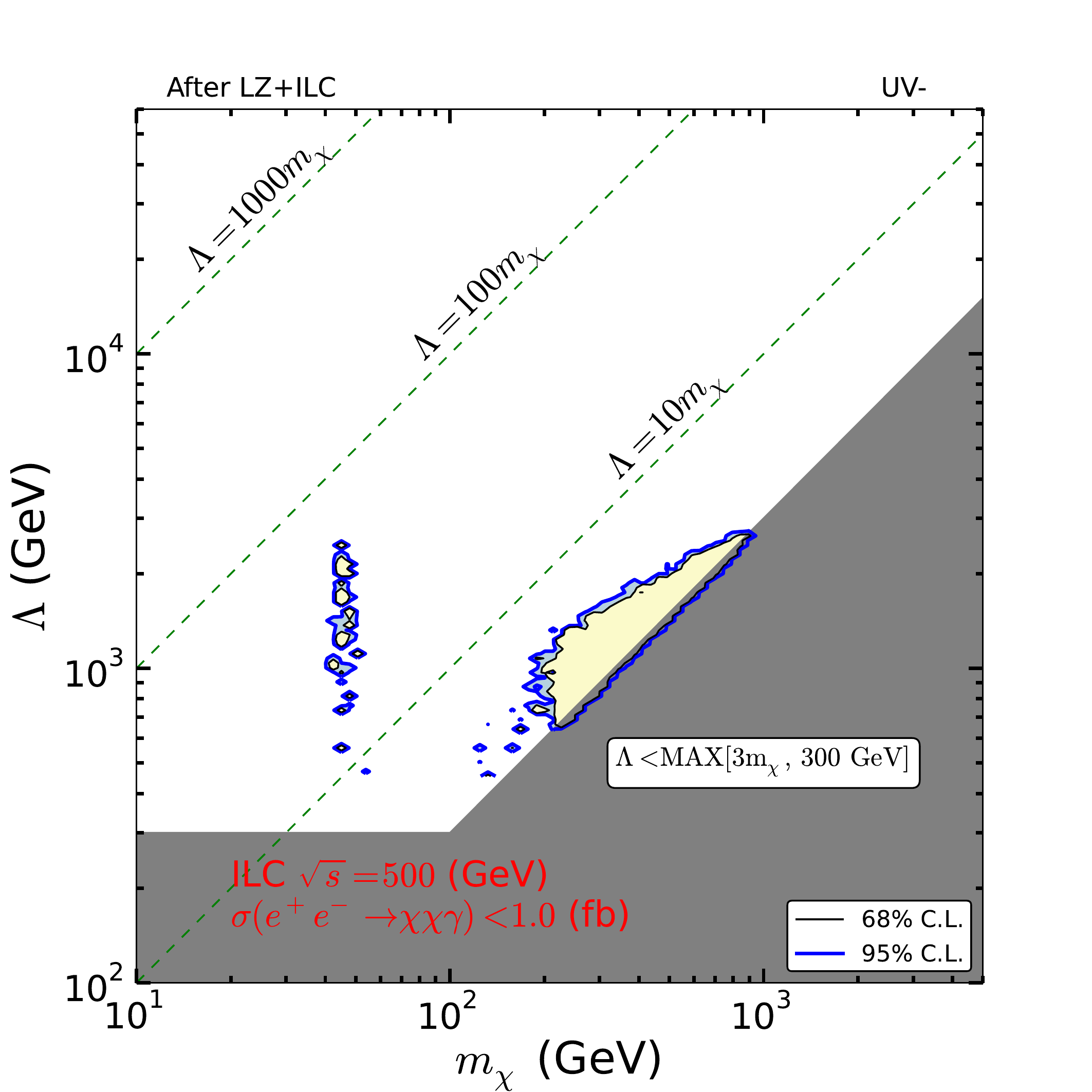}
	\caption{\sl \small Same as Fig.\,\ref{fig: mx_lam}, after imposing the expected future constraints (in the absence of a signal), from LZ and 250\,GeV ILC/CEPC for upper panels, and from LZ and 500\,GeV ILC for lower panels.}
	\label{fig: future}
\end{figure}

We show in Fig.\,\ref{fig: future} the resulting parameter space after imposing the expected future constraints, in the absence of a signal. The Higgs-pole region is found to be completely within the reach of the future SI direct detection limits, while a substantial region near the Z-pole and part of the heavier DM mass range can also be explored by the lepton colliders and SD direct detection experiments. In case a Giga-$Z$ option for the future lepton colliders or even further upgrades of the SD direct DM detections are realized, we might hope to get a handle on the remaining parameter space at the Z-pole. For dark matter masses $m_\chi \gtrsim$ 100--200\,GeV, since the four-Fermi interactions with SM quarks and leptons are mainly responsible to achieve the required DM annihilation cross-sections, the high-luminosity run of the LHC may test part of this region, while the rest can only be studied by more energetic lepton colliders such as the CLIC experiment\,\cite{Abramowicz:2013tzc}.

\vspace{0.5cm}
\noindent
{\bf Acknowledgments}
\vspace{0.1cm}\\
\noindent
This work is supported by the Grant-in-Aid for Scientific research from the Ministry of Education, Science, Sports, and Culture (MEXT), Japan and from the Japan Society for the Promotion of Science (JSPS), Nos. 26287039 and 26287039 (for S.~Matsumoto), as well as by the World Premier International Research Center Initiative (WPI), MEXT, Japan. The work of S.~Mukhopadhyay is supported in part by the U.S.~Department of Energy under grant No.~DE-FG02-95ER40896 and in part by the PITT PACC. 

\bibliographystyle{aps}
\bibliography{ref}
\end{document}